\documentclass{aa}  

\usepackage{graphicx}
\usepackage{txfonts}
\usepackage{lipsum}
\usepackage{subcaption}
\usepackage{lscape}             
\usepackage{placeins}          
\usepackage{color}
\usepackage[
    colorlinks=true,
    linkcolor=blue,
    citecolor=blue,
    urlcolor=blue
    ]{hyperref}

\begin{document}

   \title{Identification of gravitational lenses obscured by foreground light in the KiDS dataset using U-Nets and ResNets}

        \author{S. Liu\inst{1}
          \and Rui Li\inst{2}\fnmsep\thanks{Corresponding author: liruiww@gmail.com}
          \and J. Jia\inst{1}\fnmsep\thanks{Corresponding author: junjijia@whu.edu.cn}
          \and Hui Li\inst{2}
          \and Liqing Chen\inst{2}
          \and Xiaoyue Cao\inst{2}
          \and Zizhao He\inst{3,4}
          \and Valerio Busillo\inst{5}
          \and Nicola N. Napolitano\inst{6}
          \and Crescenzo Tortora\inst{5}
          \and Fucheng Zhong\inst{7}
          \and Hao Su\inst{6}
          \and Haicheng Feng\inst{7}
          \and Yue Dong\inst{8}
          \and Ran Li\inst{10}
          \and Liang Gao\inst{2,9}}
   \institute{
   $^{1}$School of physics and technology, Wuhan University, Wuhan 430072, China.\\
   $^{2}$Institute for Astrophysics, School of Physics, Zhengzhou University, Zhengzhou, 450001, China.\\
   $^{3}$Department of Physics, Nanchang University, Nanchang, 330031, China\\
   $^{4}$Center for Relativistic Astrophysics, Nanchang University, Nanchang, 330031, China\\
   $^{5}$INAF – Osservatorio Astronomico di Capodimonte, Salita Moiariello 16, I-80131, Napoli, Italy.\\
   $^{6}$Department of Physics ``E. Pancini'', University Federico II, Via Cinthia 6, 80126-I, Naples, Italy.\\
   $^{7}$School of Physics and Astronomy, Sun Yat-sen University, Zhuhai Campus, 2 Daxue Road, Xiangzhou District, Zhuhai, People’s Republic of China.\\
   $^{8}$Yunnan Observatories, Chinese Academy of Sciences, Kunming 650216, Yunnan, People's Republic of China.\\ 
   $^{9}$School of Mathematics and Physics, Xi'an Jiaotong-Liverpool University, 111 Renai Road, Suzhou, 215123, People's Republic of China.\\
   $^{10}$ School of Physics and Astronomy, Beijing Normal University,  Beijing 100875, China.}

   \date{Received May 4, 2026}
 
  \abstract{Many lensing images are often obscured by foreground light from the central galaxies, making them challenging to detect.}
   {To address the limitations of previous lens search efforts, particularly for samples with smaller $R_{\mathrm{E}}$ or faint lensed images, we developed a composite convolutional neural network framework that utilizes both U-Net and ResNet architectures for feature extraction and classification.} 
   {We propose a hybrid search method that combines U-Net and ResNet architectures to enhance the detection of the foreground galaxy-obscured lenses. Our approach consists of two main stages: first, the U-Net model separates the foreground galaxy light from potential lensing signals, creating residual images that highlight the lensing features. Next, the ResNet module performs binary classification on these residual images to detect lensing signals.}
   {We evaluated the hybrid search method with real observational data to demonstrate its effectiveness, achieving a recall of 71.5\% and a 4.5\% false positive rate at a confidence threshold of 0.6. Applying this method to over 638,398 galaxy samples from the Kilo-Degree Survey Data Release 4 and conducting thorough inspections, we identify 88 Class A, 322 Class B, and 1,758 Class C candidates.}
   {This hybrid approach significantly enhances the completeness of existing strong gravitational lensing searches and shows great potential for improving future astronomical surveys.}

   \maketitle
   \nolinenumbers

\section{Introduction}
\label{sec:introduction}

Gravitational lensing refers to the deflection of light caused by massive objects, a phenomenon predicted by general relativity (\citealt{1936Sci....84..506E}). In cases with strong gravitational lensing, this effect results in the formation of multiple images and significant arcs of light, and in optimal alignments, it can lead to the appearance of nearly complete Einstein rings. Strong lenses play a crucial role in a wide array of astrophysical and cosmological applications. They facilitate the measurement of mass distributions in galaxies and clusters without relying on stellar dynamical assumptions (e.g.,  \citealt{2009astro2010S.159K},  \citealt{2009ApJ...705.1099A},  \citealt{2019MNRAS.482..313L},  \citealt{2024OJAp....7E.120M, 2025A&A...699A.222D}), enable the investigation of dark matter by providing insights into the internal mass structure and substructure of deflecting objects (\citealt{2012Natur.481..341V, 2017MNRAS.468.1426L, Li2025ApJ...987L..31L, 2020MNRAS.492.3047H}), and help constrain cosmological parameters through lensing statistics and time-delay cosmography (\citealt{2013ApJ...766...70S} and others). Additionally, by magnifying light from distant galaxies and quasars, they effectively extend the reach of flux-limited surveys (\citealt{2023PhRvD.108l3543C},  \citealt{2023NatAs...7..959M}).

Over the past decades, wide-field imaging surveys have produced large samples of strong-lens candidates. These include the Kilo Degree Survey (KiDS; \citealt{2013ExA....35...25D}), the Dark Energy Survey (DES; \citealt{2005astro.ph.10346T}), and the Subaru Hyper Suprime-Cam lensing survey (HSC; \citealt{2012SPIE.8446E..0ZM}). Many more lenses are expected to be identified from \textit{Euclid} (\citealt{2025A&A...697A...1E}), the Vera C. Rubin Observatory (LSST; \citealt{2019ApJ...873..111I}), and the Chinese Space Station Telescope (CSST; \citealt{2025SCPMA..6880402G}) in the near future. Historically, lens identification relied on classical algorithms targeting specific morphologies (e.g., arc-finders based on image structure) or on visual inspection by experts and citizen scientists through projects such as Space Warps (\citealt{2016MNRAS.455.1171M}). While effective, these methods were not scalable to the vast amounts of data generated in modern astronomy. To address this challenge, the field has increasingly adopted machine learning, with convolutional neural networks (CNNs;  \citealt{Lecun:1998}) becoming the dominant approach due to their strength in hierarchical pattern recognition. A common and effective strategy is to train a supervised classifier—often based on established architectures such as VGG or ResNet—using simulated lens and non-lens image cutouts. The trained network then assigns a "lens score" to each real candidate, a method proven highly effective by numerous groups (e.g.,  \citealt{2017MNRAS.471..167J, 2017MNRAS.472.1129P, 2021ApJ...923...16L, 2021A&A...653L...6C}). This approach has successfully produced high-purity samples, but its effectiveness has primarily been limited to systems with large image separations and bright arcs.

Despite significant progress in current searches for gravitational lenses, the completeness of the identified lenses from these efforts remains a major concern, particularly in scientifically important areas. A primary issue is the systematic underrepresentation of small-separation lenses—those with small Einstein radii ($R_{\mathrm{E}}$), typically less than $1''$ (\citealt{2015ApJ...811...20C})—which have garnered increasing attention from the astronomical community (e.g., \citealt{2016ApJ...824...86S, 2025A&A...697A..14A} among others) with a considerable demand for exploration. For low-redshift lenses, the lensing features may end up within the central region of the foreground galaxy's light profile, making them difficult to identify. This situation is particularly challenging because the bright light from the foreground can effectively mask the faint, curved signals from background sources that typically indicate a lens. Moreover, for high-redshift lens populations, the challenges are even greater due to the nature of the angular diameter distance relation (\citealt{1933PMag...15..761E, 2007GReGr..39.1055E}). A galaxy of a given mass produces a much smaller Einstein radius at higher redshifts, further complicating detection efforts. Consequently, the identification of high-redshift lenses exhibiting large Einstein radius necessitates a correspondingly greater mass of the central galaxy. Considering the intrinsic rarity of gravitational lensing events, the occurrence of high-redshift gravitational lenses is expected to be even more infrequent. Consequently, the search for small-angle lenses effectively becomes synonymous with searching for some of the nearest or most distant lenses in the universe. Moreover, even in systems with greater angular separation between lenses and sources, faint arcs can still fall below the detection threshold, particularly when their brightness is overshadowed by intense foreground light. For ground-based surveys, these challenges are exacerbated by the effects of the point spread function (PSF). All of these selection biases create serious scientific challenges, resulting in an incomplete survey of deflector populations. Understanding the low-mass end is crucial for testing and refining models of galaxy formation and feedback, which make specific predictions about the central mass concentrations of smaller dark matter halos.

In this work, we transformed the lens-finding task from a direct classification approach to a combination of signal separation and target classification. We introduce a two-stage pipeline designed to first extract the lensing signal and then make classifications. The signal extraction is accomplished by employing a U-Net \citealt{2015arXiv150504597R}, a convolutional network with an encoder-decoder structure that has proven exceptionally effective for image-to-image translation and segmentation tasks (\citealt{2020MNRAS.497..556H, 2021ApJ...909...27H, 2023ApJS..267....2Y, 2024MNRAS.533.1426N, 2026arXiv260111054Z}). We trained the U-Net in a supervised manner to predict the smooth, symmetric light profile of the foreground galaxy from single-band image cutouts. By subtracting this prediction from the original image, we generated a residual image that effectively serves as an enhanced representation of potential lensing features. For a true lens, it should show clear arc morphology, while for a non-lens, it should present only noise or symmetric features such as spiral arms or rings. This residual image, now with a significantly enhanced arc-to-foreground contrast, was then fed to a second, standard classification network (a ResNet) to perform the final separation from lens and non-lens.

Our methodology builds on work utilizing U-Net-like architectures for lensing tasks but is fundamentally distinct in its objectives and implementation. For instance, several groups have successfully employed these U-Net-like networks for source-lens deblending, aiming to decompose a lensed image into separate images of the foreground deflector and background source \citealt{2020MNRAS.491.2481B, 2025MNRAS.543..691Z}. While effective for image separations, such networks are not optimized for the primary large-scale detection tasks. Another strategy involves using autoencoders for unsupervised anomaly detection (e.g.,  \citealt{ 2022ApJ...935....5S}). In this paradigm, a network is trained to reconstruct non-lens galaxies. When presented with a lens, it fails to reconstruct the arcs, resulting in a large residual that flags the system as an anomaly. Our approach differs from the two by adopting a targeted, supervised strategy: the U-Net is specifically trained to model and remove the foreground light. The residual image generated is not just an error map but a tailored data product where the arc signal is maximally enhanced for subsequent classification. By focusing the U-Net on signal separation rather than full reconstruction or anomaly detection, our pipeline enhances the effective signal-to-noise ratio (S/N) of faint or small-separation arcs, directly addressing the completeness problem in survey-quality data. This paper details the design, training, and application of this residual-based pipeline to data from KiDS Data Release 4 (KiDS-DR4, \citealt{2017MNRAS.472.1129P}), presenting new candidates that help bridge the gap toward the low-mass end of strong lenses.

This paper is organized as follows. In Sect. \ref{sec:data_preparation}, we detail the data used in this study and describe our strong lens simulation pipeline. Sect. \ref{sec:methodology} presents the architecture of our two-stage machine learning model. Sect. \ref{sec:performance} evaluates the performance of the networks using both simulation and real data. In Sect. \ref{sec:real-application}, we discuss the application of this trained pipeline to the KiDS-DR4 survey data and present newly identified strong lens candidates. A discussion of the model's performance and the astrophysical implications of our findings is provided in Sect. \ref{sec:discussion}. Finally, we summarize our work and present our conclusions in Sect. \ref{sec:conclusion}.

\section{Data preparation}
\label{sec:data_preparation}
The efficacy and robustness of CNNs in astronomical surveys fundamentally depend on the volume, quality, and representativeness of the training data. The confirmed population of strong gravitational lenses within the KiDS is limited to a few hundreds (\citealt{2017MNRAS.472.1129P, 2019MNRAS.484.3879P}; \citealt{2019MNRAS.482..313L, 2020ApJ...899...30L}, and others). This sample size is statistically inadequate for training deep learning models, which require at least thousands of examples to learn the complex and subtle morphological features that characterize a lensing event. Moreover, relying solely on known lenses may introduce significant selection biases into the trained model. To address these limitations and establish a robust framework for lens detection, we developed and implemented a multistage simulation pipeline. This pipeline is designed to generate a large and diverse suite of synthetic data that accurately recover the properties of the KiDS-DR4 dataset. Our methodology produces three distinct training sets used for specific machine learning tasks: \texttt{Lens\_sim}, a library of entirely simulated gravitational lens systems that includes foreground galaxies and lensing images projected into real observed background noise; \texttt{Galaxy\_sim}, which projects only the simulated foreground galaxies into real observed background noise; and \texttt{Arc\_sim}, which overlays simulated arcs onto images of actual foreground galaxies.

\subsection{Catalog and images of real data}
\label{subsec:catalog_and_images}
Our first step was to create a comprehensive catalog of potential lens galaxies from the KiDS-DR4 dataset. We began this process by selecting galaxies and calculating their $r$-band absolute magnitudes using the available multiband photometry. This calculation was performed using the spectral energy distribution (SED) fitting method through the Efficient Analysis System with YAML (EASY) from  \citealt{2018ascl.soft12006W}. Next, we utilized the well-established fundamental plane relation—an empirical correlation that links a galaxy's effective radius, mean surface brightness, and internal kinematics—to estimate the stellar velocity dispersion ($\sigma_v$) for each object. We applied a strict selection criterion (Table \ref{tb:paras}) of $\sigma_v > 200 \text{ km s}^{-1}$, a standard threshold in lens searches that effectively isolates massive early-type galaxies capable of producing strong lensing (\citealt{2008ApJ...682..964B, 2010ApJ...724..511A}). This selection resulted in a parent sample of approximately 265,266 potential lens galaxies. For this entire sample, we compiled a detailed catalog, \texttt{FG\_catalog}, containing physical parameters such as S\'ersic profiles, derived using the \texttt{GaLNets} algorithm (\citealt{2022ApJ...929..152L}), and photometric redshifts obtained via \texttt{GaZNets} \citealt{2022A&A...666A..85L}. We further refined the catalog by selecting galaxies with a S\'ersic index $n > 2.5$, an effective radius $R_{\textrm{eff}} < 5''$, and photometric redshifts $z < 0.8$, resulting in 121,713 galaxies. Concurrently, we generated 151$\times$151 pixel ($30\arcsec \times 30\arcsec$) image cutouts centered on each of these galaxies, creating a comprehensive image library referred to as \texttt{FG\_image}.

To ensure that our simulated galaxies and lens systems are embedded within a realistic background noise, we created \texttt{BG\_noise}, a library consisting of 94,520 cutouts of real observed images. These cutouts were generated by extracting 151$\times$151 pixel sections from random positions within the KiDS-DR4 images, ensuring that they did not contain bright, centrally located objects. This method guarantees that our simulated objects are overlaid onto a background reflecting the noise characteristics of the survey, including instrumental signatures, faint unresolved sources, and so on. Finally, to model the lensed background objects, we selected a population of high-redshift source galaxies from the COSMOS DC2 (\citealt{2019ApJS..245...26K}) simulation catalog (\texttt{SR\_catalog}). This catalog offers a physically realistic distribution of galaxy colors and redshifts that are suitable for the source-plane population in a lensing survey.

\subsection{Lens and galaxy simulations}
\label{subsec:lens_galaxy_simulation}
With these foundational datasets established, we now describe our simulation pipeline. The first fully synthetic dataset, \texttt{Galaxy\_sim}, containing 180,000 images in total, was created to represent normal galaxies for training the U-Net. For each image, we randomly selected a set of S\'ersic parameters from our \texttt{FG\_catalog}. These parameters were used to create an idealized galaxy light profile, which was then convolved with a realistic PSF sampled from the KiDS survey (e.g.,  \citealt{2018MNRAS.480.1057R}). We then added the Poisson noise and finally embedded the resulting galaxy into a randomly chosen background patch from our \texttt{BG\_noise} library. The second dataset, \texttt{Lens\_sim}, comprises 18,000 fully simulated strong gravitational lens systems. In this dataset, the light distribution of each foreground lens galaxy was modeled using S\'ersic parameters randomly drawn from \texttt{FG\_catalog}. The corresponding mass distribution was described by a singular isothermal ellipsoid (SIE) model (\citealt{1994A&A...284..285K}), which is widely adopted as an effective parameterization for massive elliptical galaxies. To ensure physical consistency, we required the ellipticity and position angle of the mass distribution to be the same as those of the light profile. The Einstein radius ($R_{\textrm{E}}$) was calculated based on the velocity dispersion of the lens and the angular diameter distances to the lens ($D_L$) and source ($D_S$). A background source drawn from \texttt{SR\_catalog} was placed in the source plane, and its light was ray-traced through the gravitational potential of the SIE lens to produce the lensed images. This complete lensed system underwent the same post-processing as the \texttt{Galaxy\_sim} images, including convolution with the KiDS PSF, addition of Poisson noise, and embedding into a real background from \texttt{BG\_noise}. A third dataset, \texttt{Arc\_sim}, was created by incorporating simulated arcs into real foreground galaxies (\texttt{FG\_image}). In this process, we selected a real galaxy directly from the \texttt{FG\_image} library, and calculated its lensing potential using its corresponding physical parameters from \texttt{FG\_catalog} (e.g., $\sigma_v$ and redshift) by assuming the SIE mass profile. We then simulated only the lensed arcs from a background source and directly injected these lensed images into the real image of the foreground galaxy. The primary advantage of this method is its ability to preserve the authentic, complex morphology, local environment, and noise characteristics of the true foreground galaxy, which are difficult to replicate in purely simulated images. 

\begin{table}[ht]
\caption{\label{tb:paras}Range and distribution of parameters in our lens simulation.}
\begin{tabular}{cccc}
\hline \hline
\noalign{\smallskip}
Parameter&Range&Units&Distribution\\
\noalign{\smallskip}
\hline
\noalign{\smallskip}
\multicolumn{4}{c}{lens (SIE)} \\
\noalign{\smallskip}
\hline
\noalign{\smallskip}
Einstein radius    &0.75-3  &arcsec& ... \\
Velocity dispersion&200-800 &km/s  & real \\
External shear     &0-0.1   &...   & uniform \\
\noalign{\smallskip}
\hline
\noalign{\smallskip}
\multicolumn{4}{c}{Source (S\'ersic)} \\
\noalign{\smallskip}
\hline
\noalign{\smallskip}
Effective radius   &0.05-0.5&arcsec& normal \\
Axis ratio         &0.3-1.0 &...   & uniform \\
S\'ersic index     &0.3-8.0 &...   & uniform \\
\noalign{\smallskip}
\hline
\noalign{\smallskip}
\multicolumn{4}{c}{Foreground (S\'ersic)} \\
\noalign{\smallskip}
\hline
\noalign{\smallskip}
Velocity dispersion& $>$200 &km/s& real \\
Effective radius   & $<$5   &arcsec& real \\
S\'ersic index     & $>$2.5 & ...  & real \\
Redshift           & $<$0.8 & ...  & real \\
Magnitude         & $<$20  & ...  & real \\
\noalign{\smallskip}
\hline \hline
\end{tabular}
\tablefoot{We extract the foreground galaxy parameters directly from \texttt{FG\_catalog} in \texttt{Lens\_sim}, with their distributions following the input catalog. We calculate the Einstein radius using the velocity dispersion alongside the foreground and background redshifts. The foreground (S\'ersic) parameter ranges listed in the table represent the selection criteria for our catalog.}
\end{table}

This simulation process produces three datasets for two distinct machine learning tasks. The purely synthetic datasets, \texttt{Galaxy\_sim} (non-lenses) and \texttt{Lens\_sim} (lenses), provide perfect ground truth, making them ideal for training a U-Net model for the segmentation of lensed features. For the final classification task, we employed a ResNet model, with the \texttt{Arc\_sim} images serving as the positive training class ("lenses"). To construct the negative training class, we randomly selected 170,000 galaxies from our target search sample while carefully excluding all previously known and candidate lenses. This strategy makes the classifier more effectively trained to distinguish lenses from other types of galaxies. Overall, this approach provides a robust training sample, crucial for developing a high-performance lens detection pipeline optimized for the KiDS survey.

   \begin{figure*}[htbp]
        \centering
        \includegraphics[width=2.7cm]{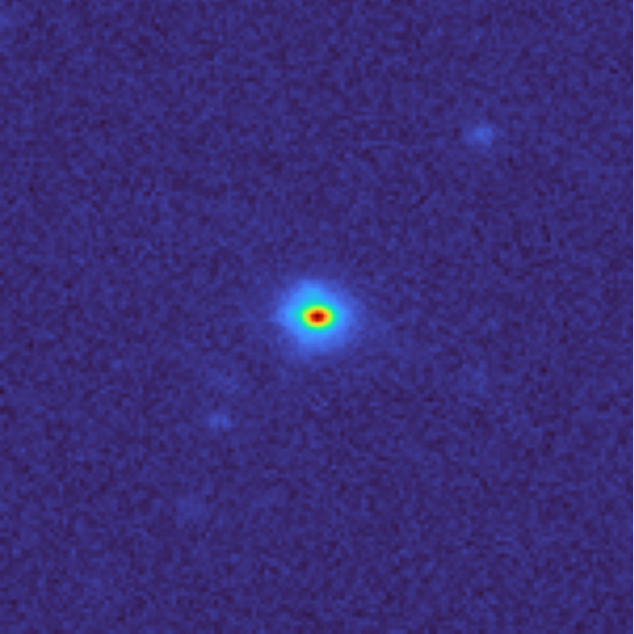}
        \includegraphics[width=2.7cm]{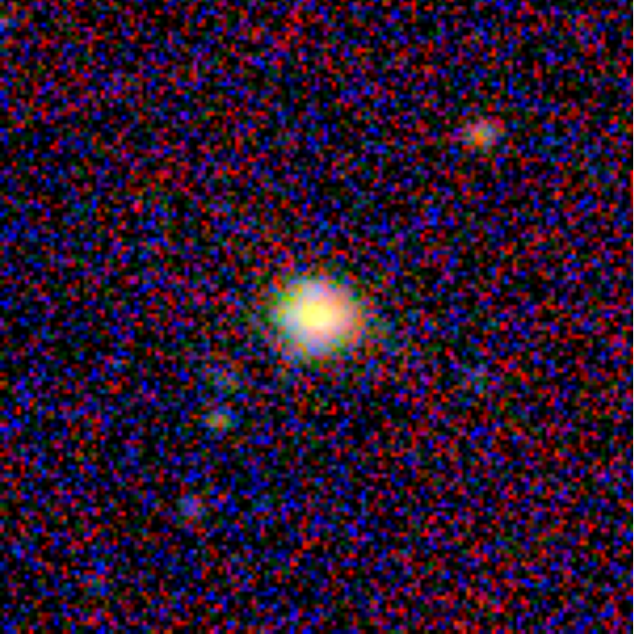}
        \quad
        \includegraphics[width=2.7cm]{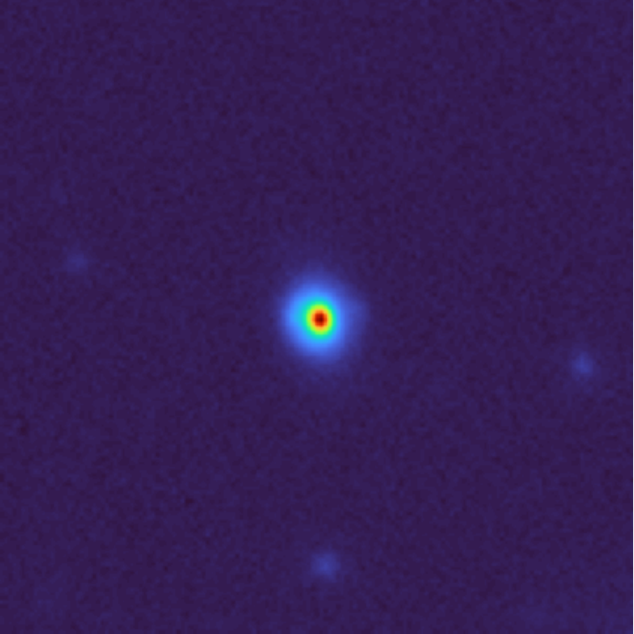}
        \includegraphics[width=2.7cm]{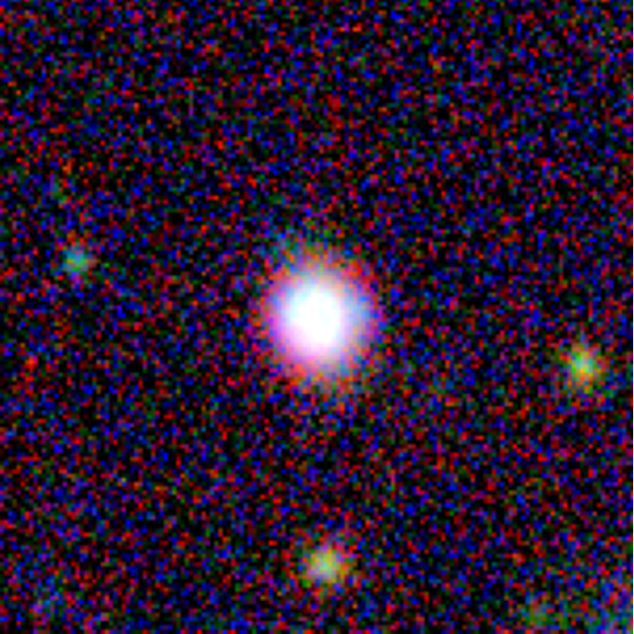}
        \quad
        \includegraphics[width=2.7cm]{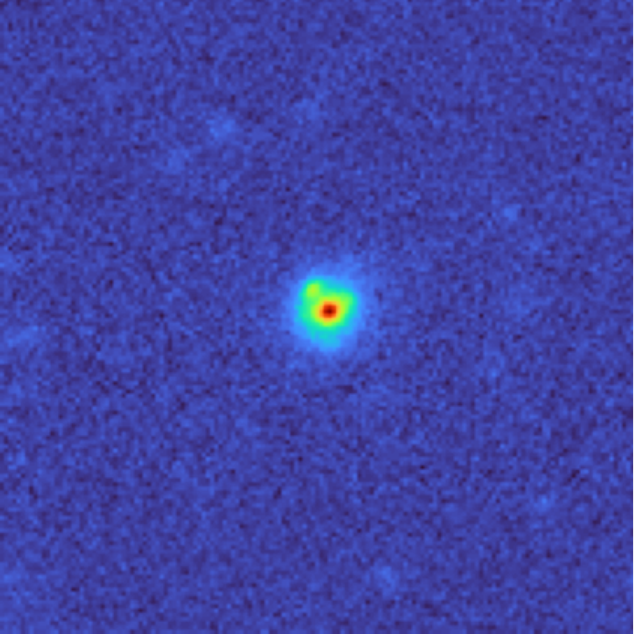}
        \includegraphics[width=2.7cm]{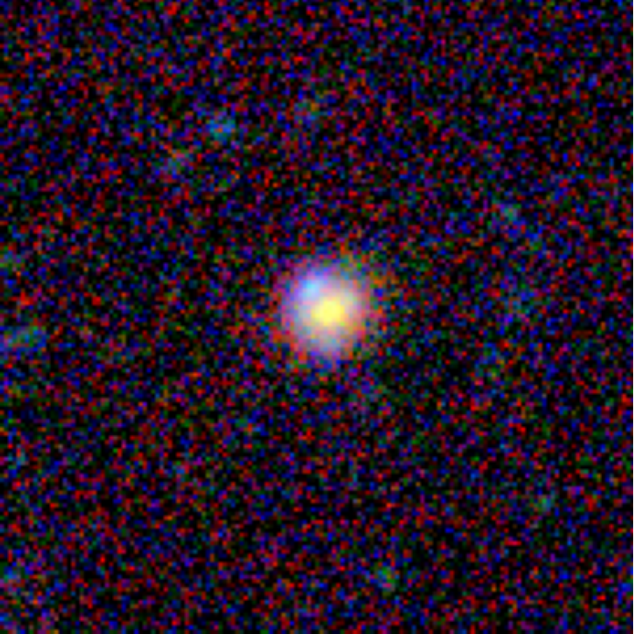}
        \newline
        \vspace{-5mm}
        
        \includegraphics[width=2.7cm]{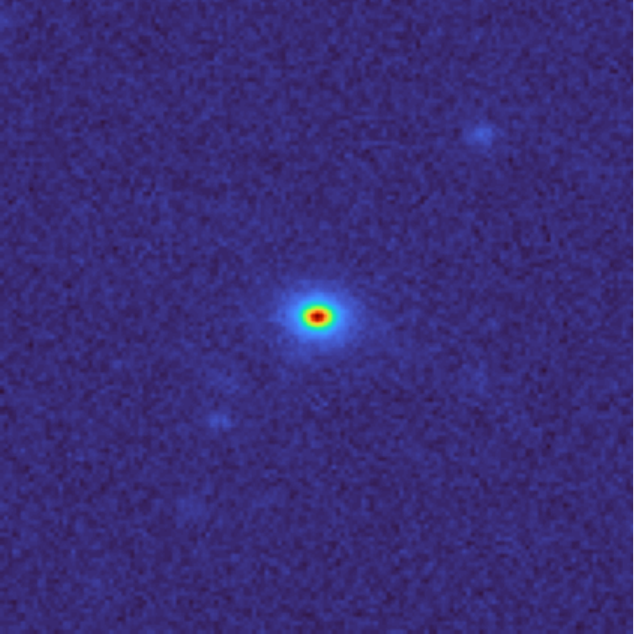}
        \includegraphics[width=2.7cm]{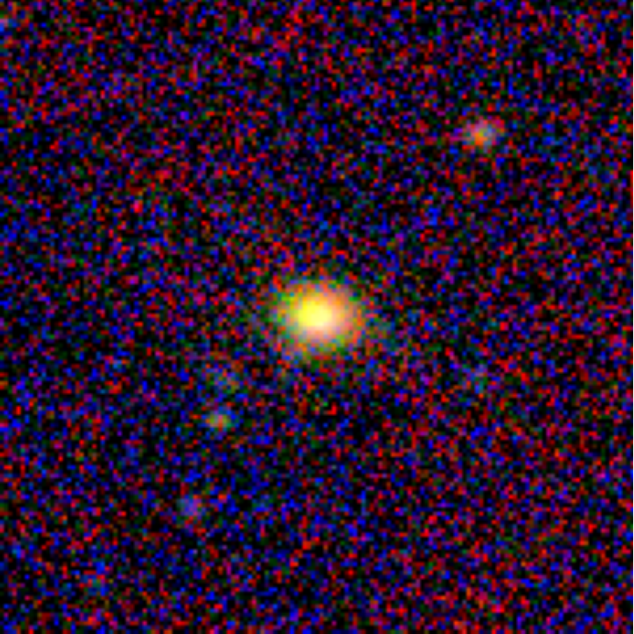}
        \quad
        \includegraphics[width=2.7cm]{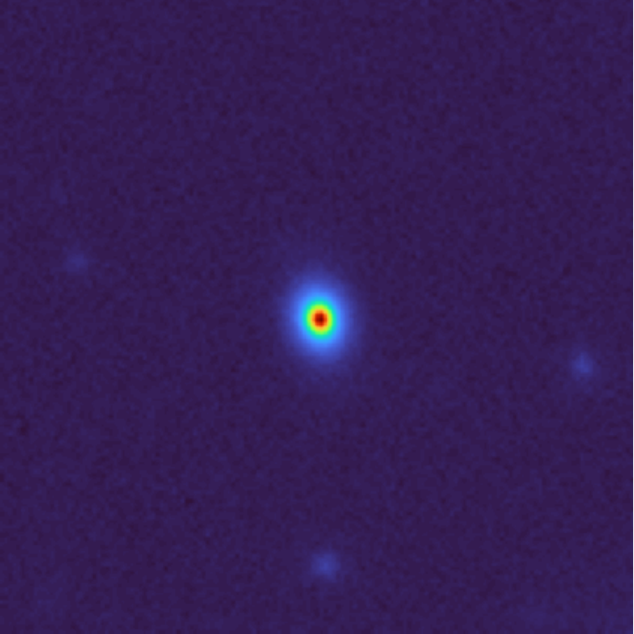}
        \includegraphics[width=2.7cm]{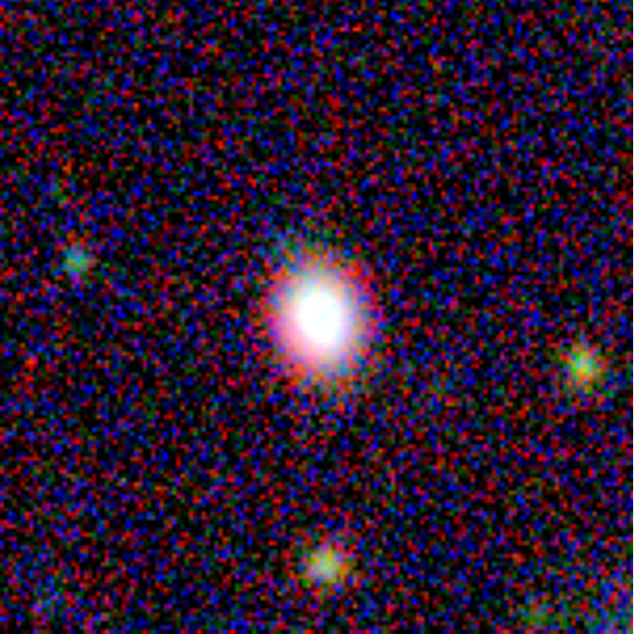}
        \quad
        \includegraphics[width=2.7cm]{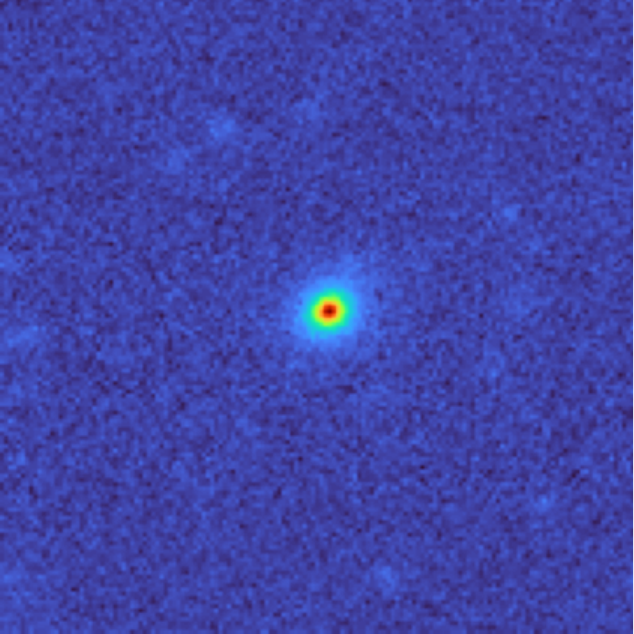}
        \includegraphics[width=2.7cm]{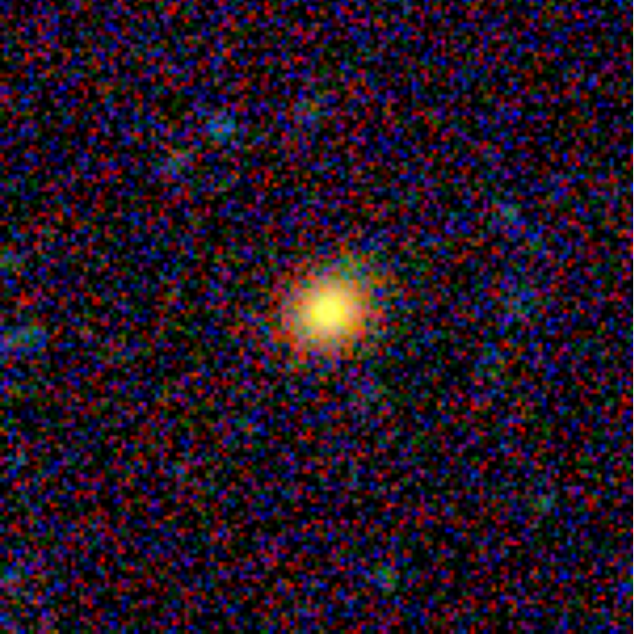}
        \newline
        \vspace{-5mm}
        
        \includegraphics[width=2.7cm]{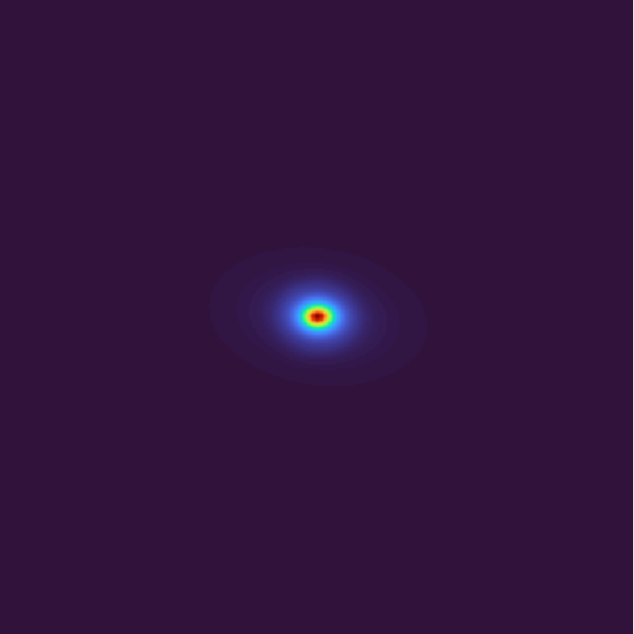}
        \includegraphics[width=2.7cm]{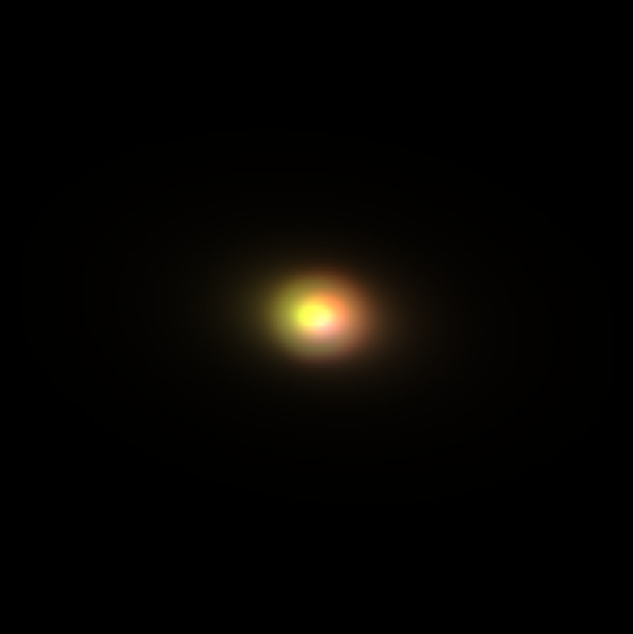}
        \quad
        \includegraphics[width=2.7cm]{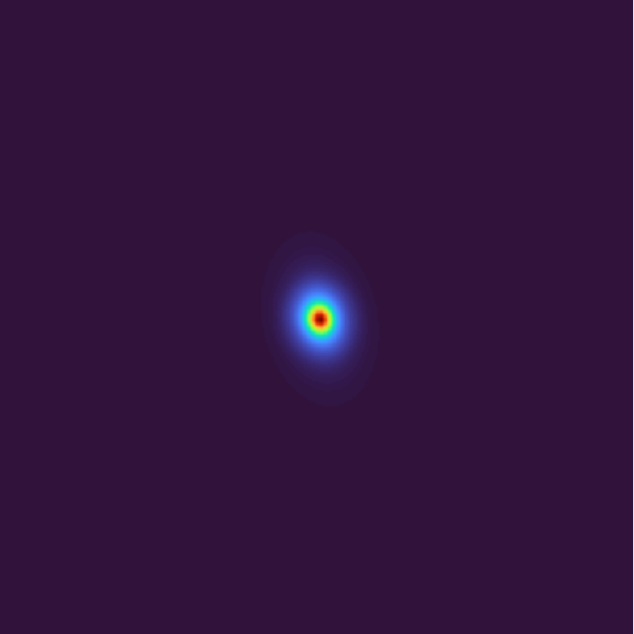}
        \includegraphics[width=2.7cm]{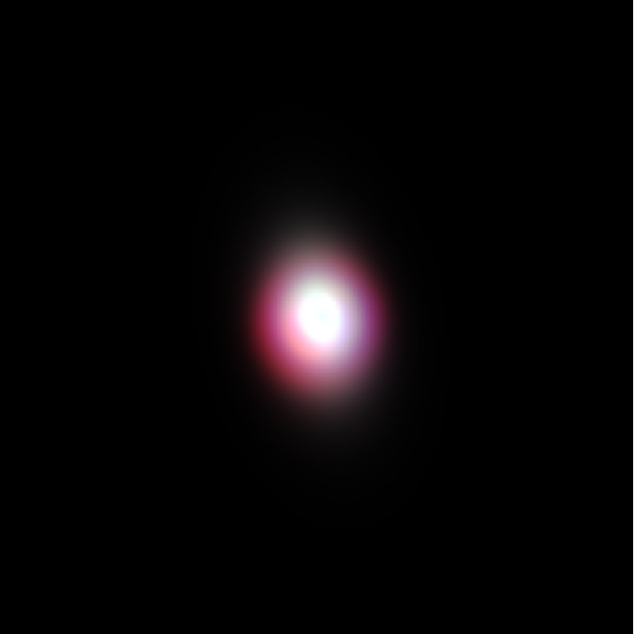}
        \quad
        \includegraphics[width=2.7cm]{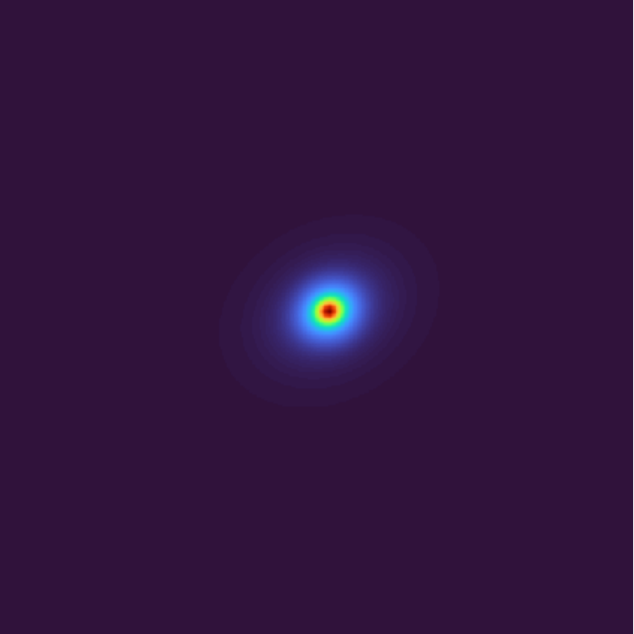}
        \includegraphics[width=2.7cm]{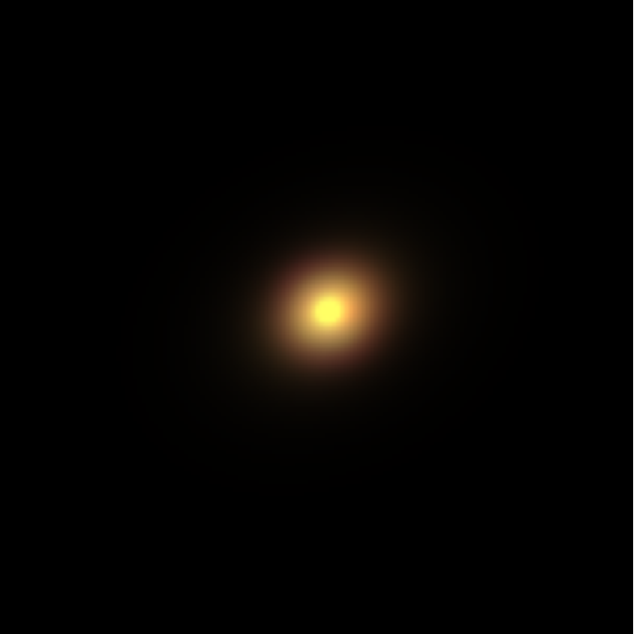}
        \newline
        \vspace{-5mm}
        
        \includegraphics[width=2.7cm]{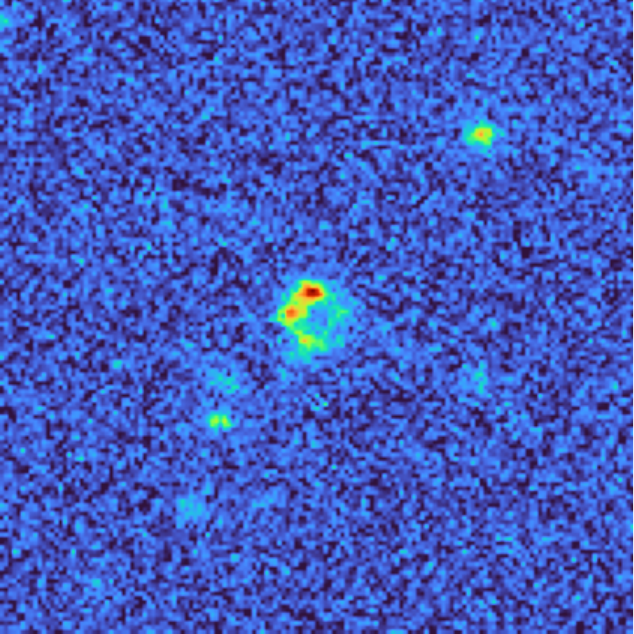}
        \includegraphics[width=2.7cm]{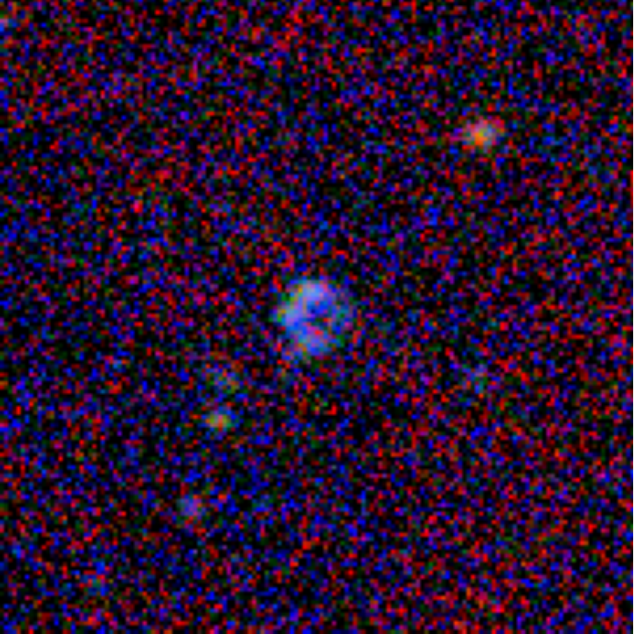}
        \quad
        \includegraphics[width=2.7cm]{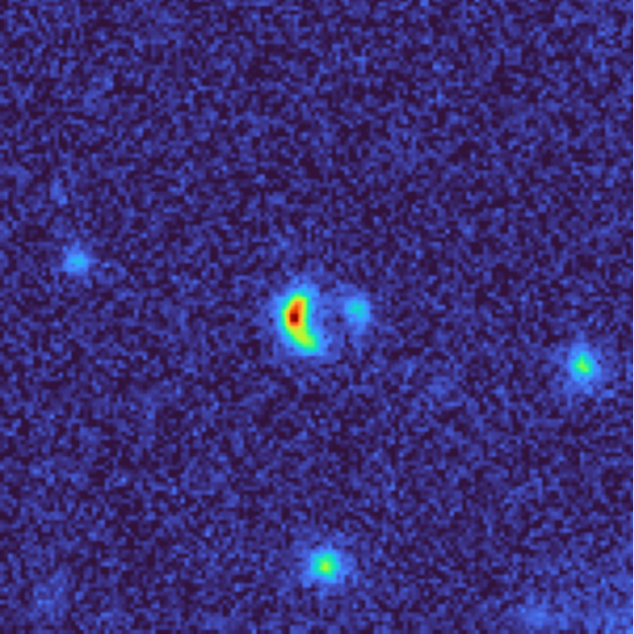}
        \includegraphics[width=2.7cm]{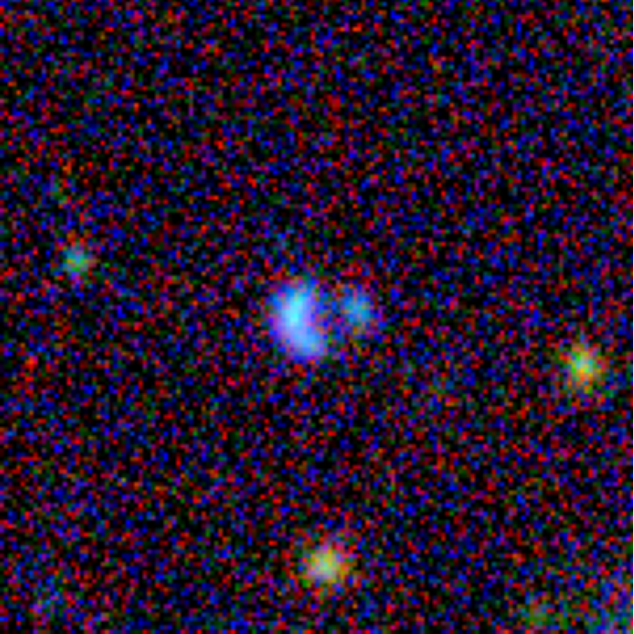}
        \quad
        \includegraphics[width=2.7cm]{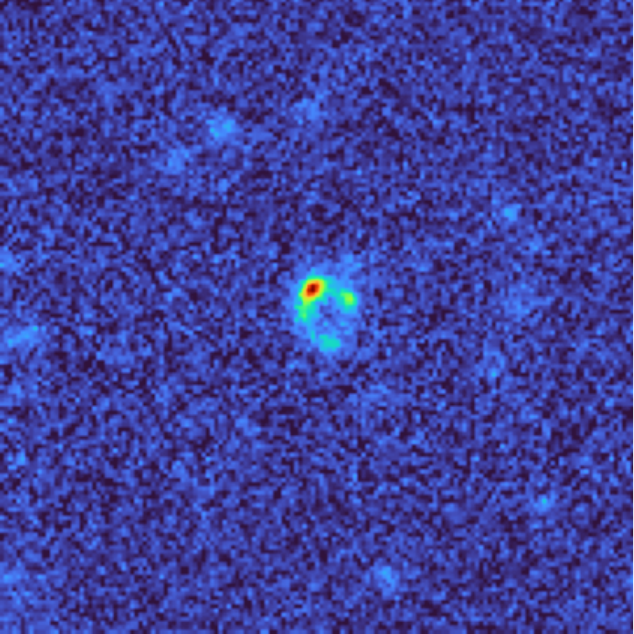}
        \includegraphics[width=2.7cm]{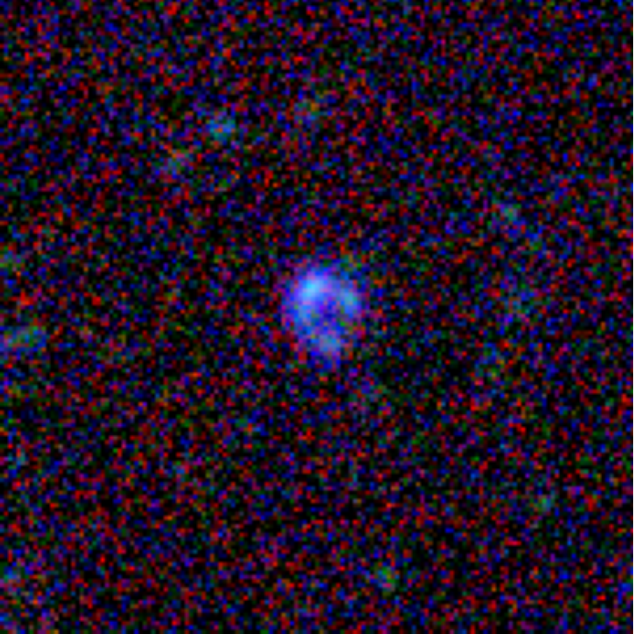}
        \newline
        \vspace{-5mm}
        \caption*{(a)}

        \includegraphics[width=2.7cm]{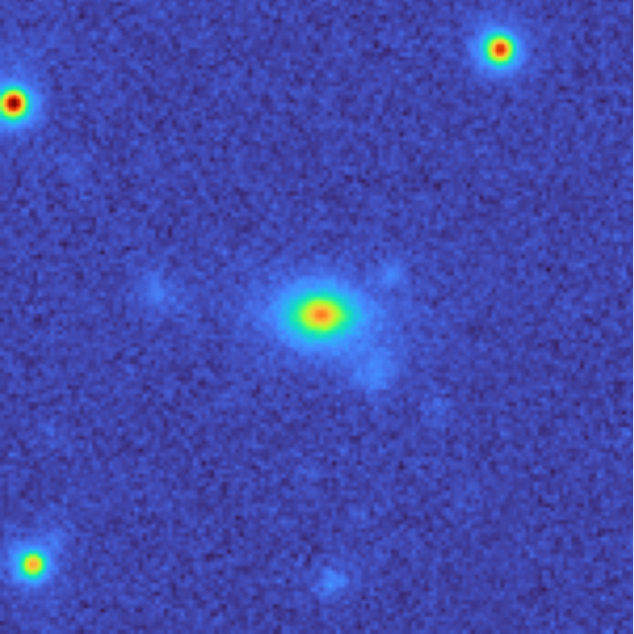}
        \includegraphics[width=2.7cm]{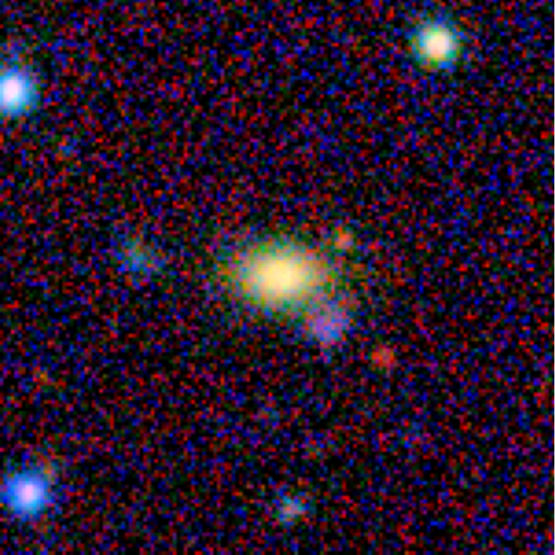}
        \quad
        \includegraphics[width=2.7cm]{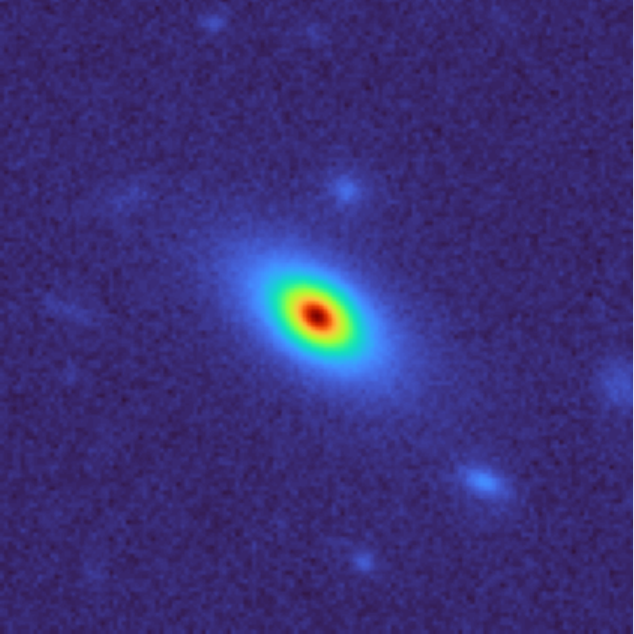}
        \includegraphics[width=2.7cm]{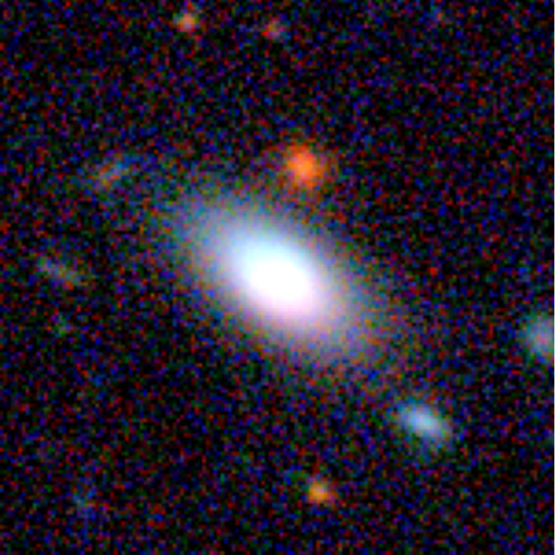}
        \quad
        \includegraphics[width=2.7cm]{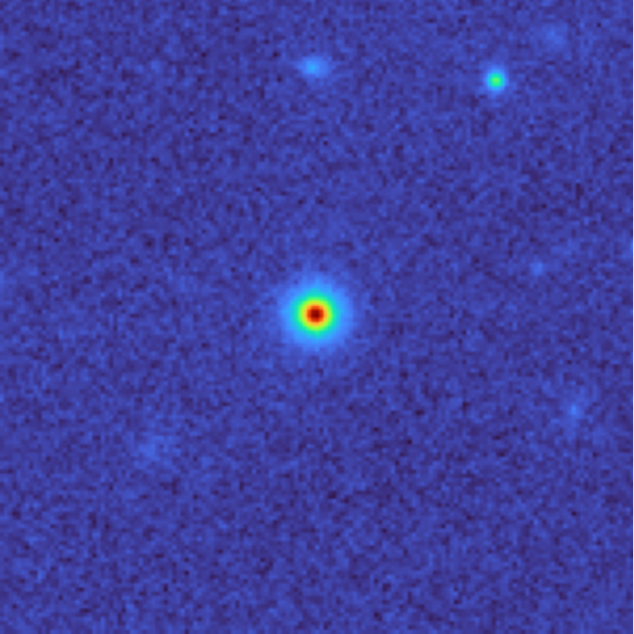}
        \includegraphics[width=2.7cm]{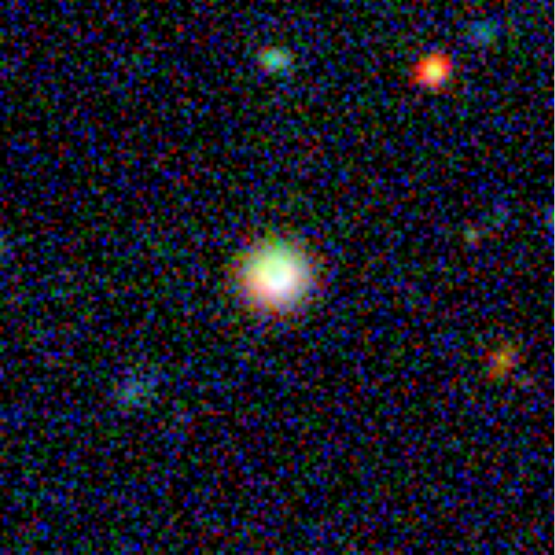}
        \newline
        \vspace{-5mm}

        \includegraphics[width=2.7cm]{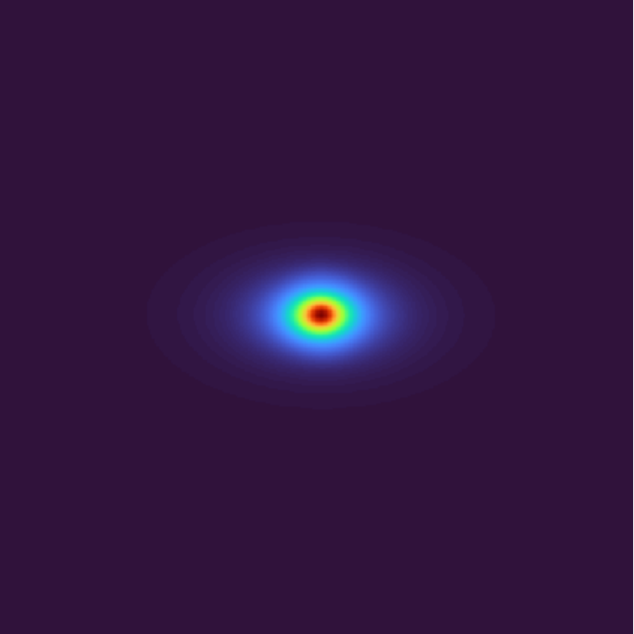}
        \includegraphics[width=2.7cm]{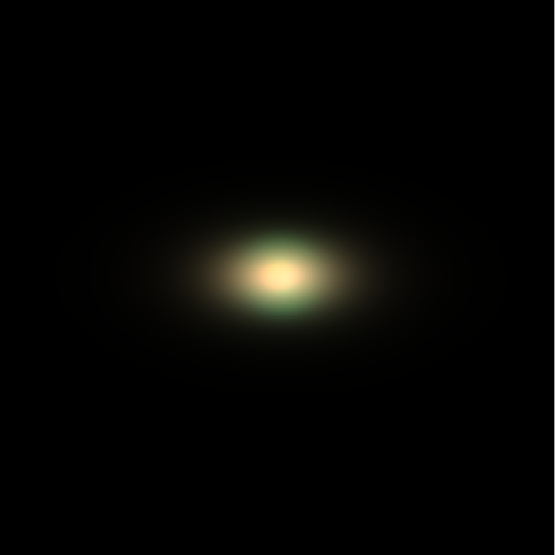}
        \quad
        \includegraphics[width=2.7cm]{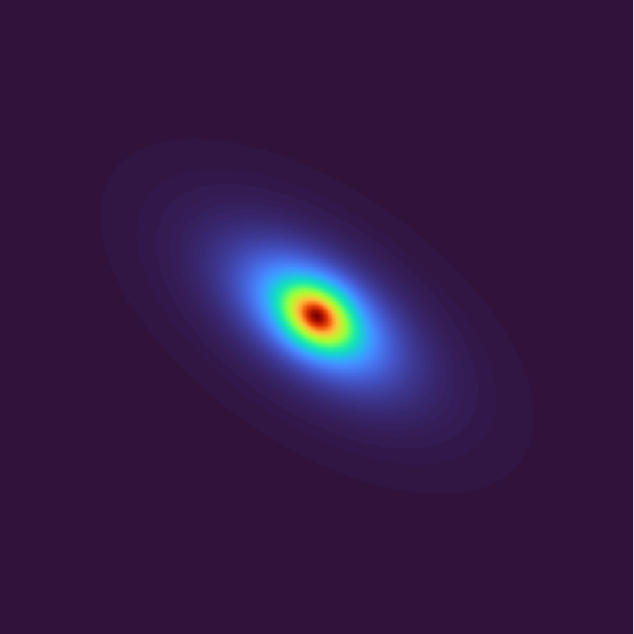}
        \includegraphics[width=2.7cm]{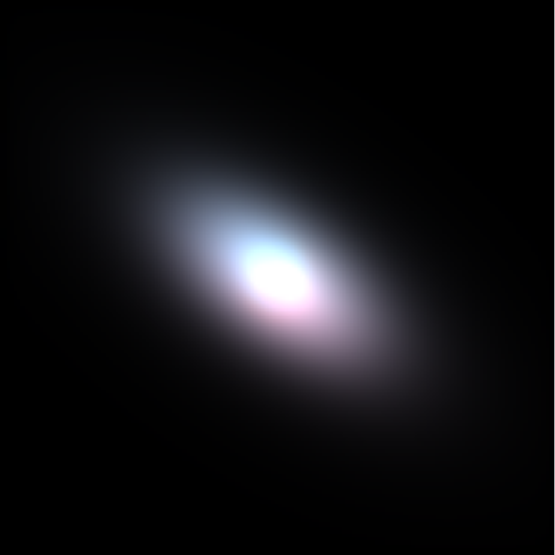}
        \quad
        \includegraphics[width=2.7cm]{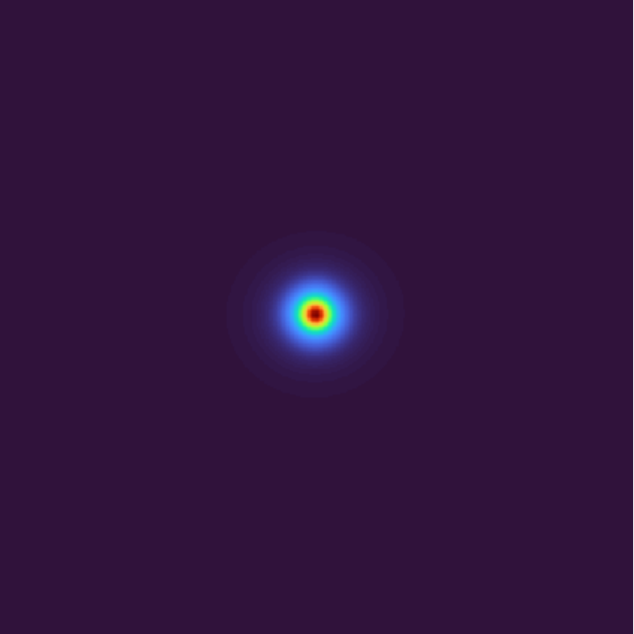}
        \includegraphics[width=2.7cm]{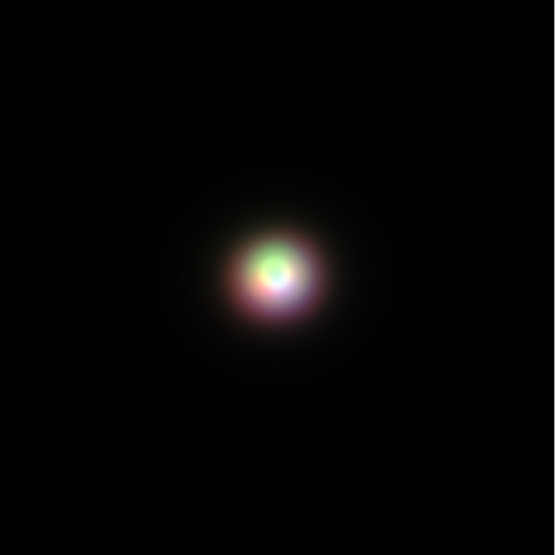}
        \newline
        \vspace{-5mm}
        \caption*{(b)}
        
        \caption{$r$-band and colored simulation images. Subfigure (a): Three lenses. Subfigure (b): Three non-lenses. Each pair of columns represents one group image of an independent sample. Odd-number column: $r$-band grayscale images. Even-number column: Color-composite image of the $g$, $r$, and $i$ bands. Subfigure (a), top to bottom: Whole lens system, the central galaxy with noise, the central galaxy without noise, and the lens arc with noise. Subfigure (b), top to bottom: Central galaxy with noise and central galaxy without noise.}
        \label{fig:sim}
    \end{figure*}

Compared to our previous work (\citealt{2020ApJ...899...30L, 2021ApJ...923...16L}), the new simulation introduces several key improvements. First, prior methodologies often rely on either a uniform PSF for all cutouts (e.g.,  \citealt{2019PhDT.......108P},  \citealt{2020ApJ...899...30L}) or randomly simulated PSFs for each cutout (e.g.,  \citealt{2021ApJ...923...16L}). In contrast, in this work we utilized a specific Shapelet PSF model (\citealt{Kuijken_2019}) corresponding to the sky coordinates of the background patch into which the simulated object is injected. This approach ensures a spatially variable PSF that accurately reflects the observational conditions and improves the authenticity and reliability of the simulation results. Furthermore, we adopted a new method for calculating the S/N of the lensed arcs, following the framework established by  \citealt{2015ApJ...811...20C}. Instead of sampling only a few pixels near the brightest point of an arc, our new calculation first identifies all pixels belonging to each lensed image. The S/N is then computed using the integrated flux from all these pixels. This method ensures that the S/Ns of the lensed features are calculated robustly, allowing us to retain those pixels above the background noise while removing the fainter ones. This new methodology significantly benefits model training by enhancing the quality of the positive training set and minimizing contamination from lens systems with unclear features, thereby ensuring the quality of the lensing images learned by models to avoid a certain degree of data contamination. In this work, we required the arc component of an image to have a S/N exceeding 20 to be included in the final training set. This threshold represents a carefully chosen balance: it is low enough to train the model on features that may be obscured by foreground light, yet high enough to ensure that most simulated arcs are visually detectable if looked at by humans (see Sect. \ref{subsec:visual_classification}). While this strategy maximizes our potential to identify candidates with small $R_{\mathrm{E}}$ or faint lensed images, we acknowledge the inherent risks. Some low-S/N images may contaminate the training data and slightly degrade the model's overall classification performance. Achieving the optimal balance between discovery efficiency and model accuracy is a significant challenge.

\subsection{Training and testing data}
\label{subsec:traing_testing_data}
All generated images were set to a uniform size of 151 $\times$151 pixels, corresponding to a field of view of $30\arcsec \times 30\arcsec$. The simulation pipeline is capable of producing multiband ($g$, $r$, and $i$) color-composite images for each simulated system, as well as separate image planes for each component (central galaxy, lensed arc, background) in each band, as illustrated in Fig. \ref{fig:sim}. In this work, we utilized only the $r$-band. While neural networks often exhibit enhanced performance with multiband color data during the training process, they still face significant challenges when applied to real data, as demonstrated by  \citealt{2021ApJ...923...16L}. Therefore, we leave the implementation of multiband inputs for future work.

The simulated data were categorized into distinct datasets for training and testing the U-Net and ResNet models. The training dataset was divided into two categories. The first category, \texttt{Train\_unet}, was used to train the U-Net to obtain the residuals. It comprises 18,000 simulated lens images from the \texttt{Lens\_sim} dataset and 180,000 simulated non-lens images from the \texttt{Galaxy\_sim} dataset. We address the issue of unbalanced data distribution in the Sect. \ref{sec:discussion}. The second category, \texttt{Train\_res}, includes 170,000 lenses from the \texttt{Arc\_sim} and 170,000 galaxy images with different types of morphologies from the \texttt{FG\_catalog} library; it was used to train the ResNet model (Fig. \ref{fig:process}). Both the U-Net and ResNet share the same testing dataset, \texttt{Test\_sim}, which consists of 4,000 lenses generated by \texttt{Lens\_sim} and 4,000 images from the \texttt{Galaxy\_sim}. To ensure performance on real data, we also collected another test set, named \texttt{Test\_real}, which contains 200 real lens images randomly collected from  \citealt{2019MNRAS.484.3879P},  \citealt{2020ApJ...899...30L}, and  \citealt{2021ApJ...923...16L}, as well as 200 non-lens galaxy images sourced from the \texttt{FG\_catalog} library. The 200 positive samples consist exclusively of high-probability lens candidates identified in previous studies and have undergone manual visual inspection by experts.
Notably, we made no overlap between the images across the datasets; each data instance appears exclusively in one dataset. The dataset partitioning has been summarized and is presented in Table \ref{tb:dataset}.

\begin{table}[ht]
\caption{\label{tb:dataset}Dataset partitioning list.}
\small
\begin{tabular}{ccc}
\hline \hline
\noalign{\smallskip}
Dataset&Data quantity&Data source\\
\noalign{\smallskip}
\hline
\noalign{\smallskip}
\texttt{Train\_unet}&18,000+180,000&\texttt{Lens\_sim}+\texttt{Galaxy\_sim}\\
\texttt{Train\_res}&170,000+170,000&\texttt{Arc\_sim}+\texttt{FG\_catalog} \\
\texttt{Test\_sim}&4,000+4,000&\texttt{Lens\_sim}+\texttt{Galaxy\_sim} \\
\texttt{Test\_real}&200+200&\texttt{Real\ lenses}+\texttt{FG\_catalog}\\
\noalign{\smallskip}
\hline \hline
\end{tabular}
\tablefoot{The "Data quantity" column represents the number of positive and negative samples. The "Data source" column indicates the generation method or source of the data. \texttt{Real\ lenses} represents the real observation lenses.}
\end{table}

\section{CNN-based search methodology}
\label{sec:methodology}
Our strong lensing search methodology consists of a two-stage process that leverages two different CNN architectures: U-Net and ResNet. The core idea is to transform the challenging task of identifying faint lensed arcs within a bright galactic foreground into a more manageable problem of signal detection within a noise-like residual image. The workflow is as follows. First, we trained a U-Net model not to find arcs, but to reconstruct and effectively remove the light from the foreground galaxy. By subtracting this reconstructed foreground from the original image, we generated a residual image. This residual either contains only the lensed arc features and background noise (if a lens is present), or just noise (if it is not). Second, this residual image was passed to a ResNet classifier specifically trained to determine if the residual contains the signal of lensed arcs. This entire pipeline was implemented using the Keras\footnote{\url{https://keras.io}} API with a TensorFlow\footnote{\url{https://www.tensorflow.org}} backend.

\subsection{Structures of the U-Net and ResNet}
\label{sec:resnet}
\begin{figure*}[ht!]
    \centering
    {\includegraphics[width=18cm]{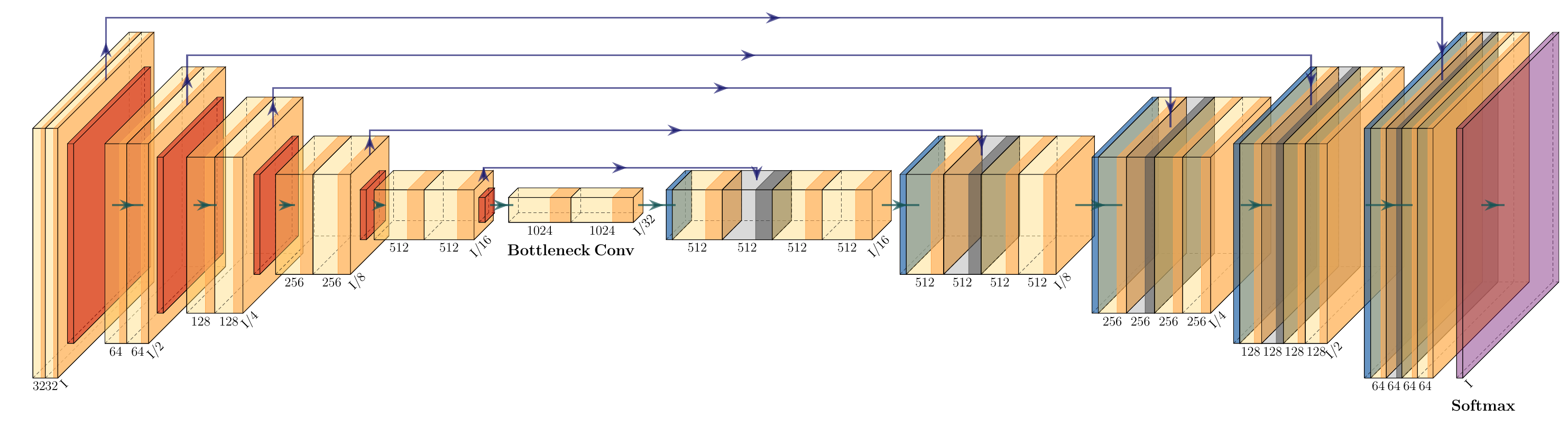}}
     \caption{Construction of the U-Net model.  The input and output size of the
network is ($96\times96\times1$), as well as the output. Every kernel in the convolutional layer and deconvolutional layer has a size of $3\times3$. The kernel size in the pooling layer and up-pooling layer is $2\times2$. The network ends with a global average pooling layer.}
     \label{fig:unet}
\end{figure*}

\begin{figure}[ht!]
\centering
\includegraphics[width=\hsize]{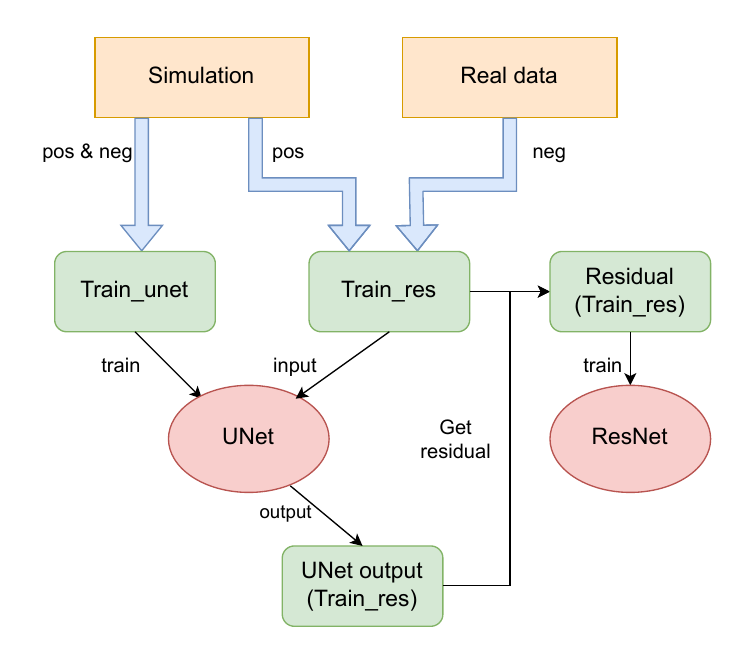}
    \caption{Full training process map. The yellow rectangles represent the data sources from either the simulation algorithm or real observations. The green rectangles represent the dataset, while the red rectangles represent the CNN models.}
    \label{fig:process}
\end{figure}

\begin{figure*}
    \centering
    \includegraphics[width=0.48 \textwidth]{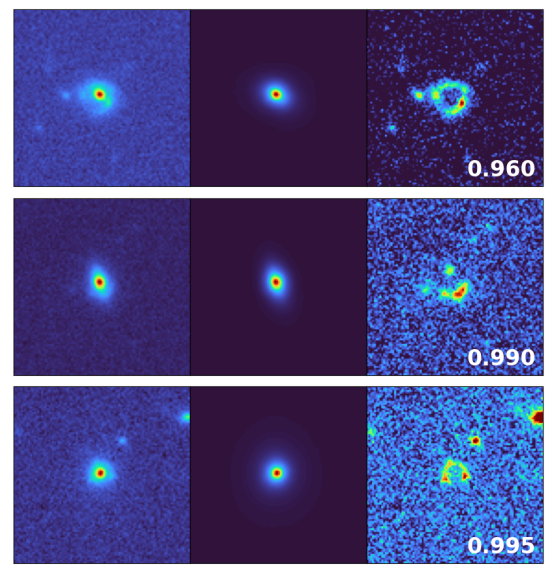}
    \includegraphics[width=0.48 \textwidth]{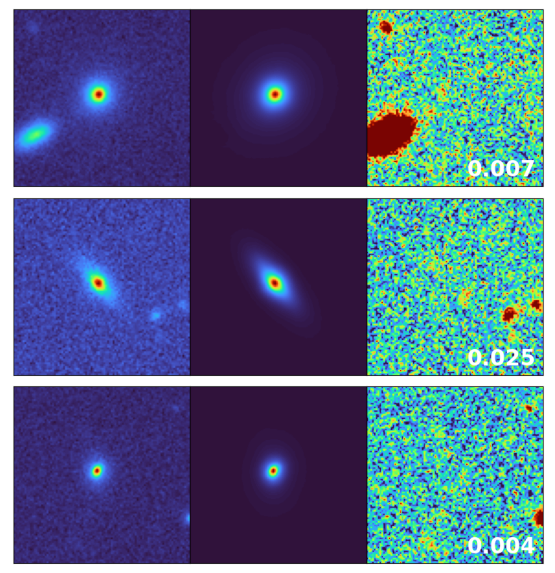}
    \caption{U-Net predictions on \texttt{Test\_real}. Additional cases are shown in Fig.~\ref{fig:u_pred_real_apd}. These are a few examples of \texttt{Test\_real} predictions. Left: Lens results. Right: Non-lens results. For each subgraph, the left column shows the original figure; the middle column shows the UNet output videlicet the central galaxy without noise; and the right column shows the residual of the first two, the lens arc or the background noise. Their image sizes are all $96\times96$ pixels. The number is the probability given by the ResNet ($P_{\textrm{ResNet}}$).}
    \label{fig:u_pred_real}
\end{figure*}

The U-Net architecture was originally developed for biomedical image segmentation, where it achieved state-of-the-art results \citealt{2015arXiv150504597R}. Its name comes from its characteristic U-shaped structure, which consists of a contracting path (encoder) that captures context and a symmetric expanding path (decoder) enabling precise localization. The use of skip connections between the encoder and decoder enables the network to combine deep semantic features with shallow, fine-grained details, making it highly effective for segmentation tasks. While other object detection algorithms, such as Mask R-CNN \citealt{2017arXiv170306870H}, exist, U-Net's lightweight architecture enhances its efficiency \citealt{2022ISD-52-383}. Its success has led to widespread adoption in astronomy for tasks such as sunspot segmentation \citealt{2025ApJ...980..261C} and reionization signal extraction \citealt{2025ApJ...988...84G}. Although U-Net has been used in previous lens searches \citealt{2024MNRAS.533.1426N}, we employed it here in a novel way for foreground subtraction. In this work, we constructed an 11-layer U-Net, as shown in Fig. \ref{fig:unet}\footnote{The program for drawing the neural network structure diagram is indebted to \url{https://github.com/HarisIqbal88/PlotNeuralNet/tree/master}.}. The network is designed to accept a single-channel input image of size 96 $\times$ 96 and produce an output of the same dimensions. A key aspect of our training strategy is the definition of the network's target. Instead of training the U-Net to predict the lensed arc, we trained it to predict the light profile of the central galaxy (see Fig. \ref{fig:u_pred_real}). This approach defines the input and output of the U-Net during the training process (Fig. \ref{fig:process}). Once trained, the U-Net model serves as a predictor for the foreground galaxy. When we input an image from our dataset into the network, it output a clean model of the central galaxy. We then generated a residual image by subtracting this U-Net output from the original input image. As illustrated in Fig. \ref{fig:u_pred_real}, if the input image contains a gravitational lens, the resulting residual primarily consists of the lensed arc features along with background noise. If the input is a non-lens, the residual contains only background noise. These residual images form the input for the second stage of our pipeline.

\begin{figure}[ht!]
\centering
\includegraphics[width=\hsize]{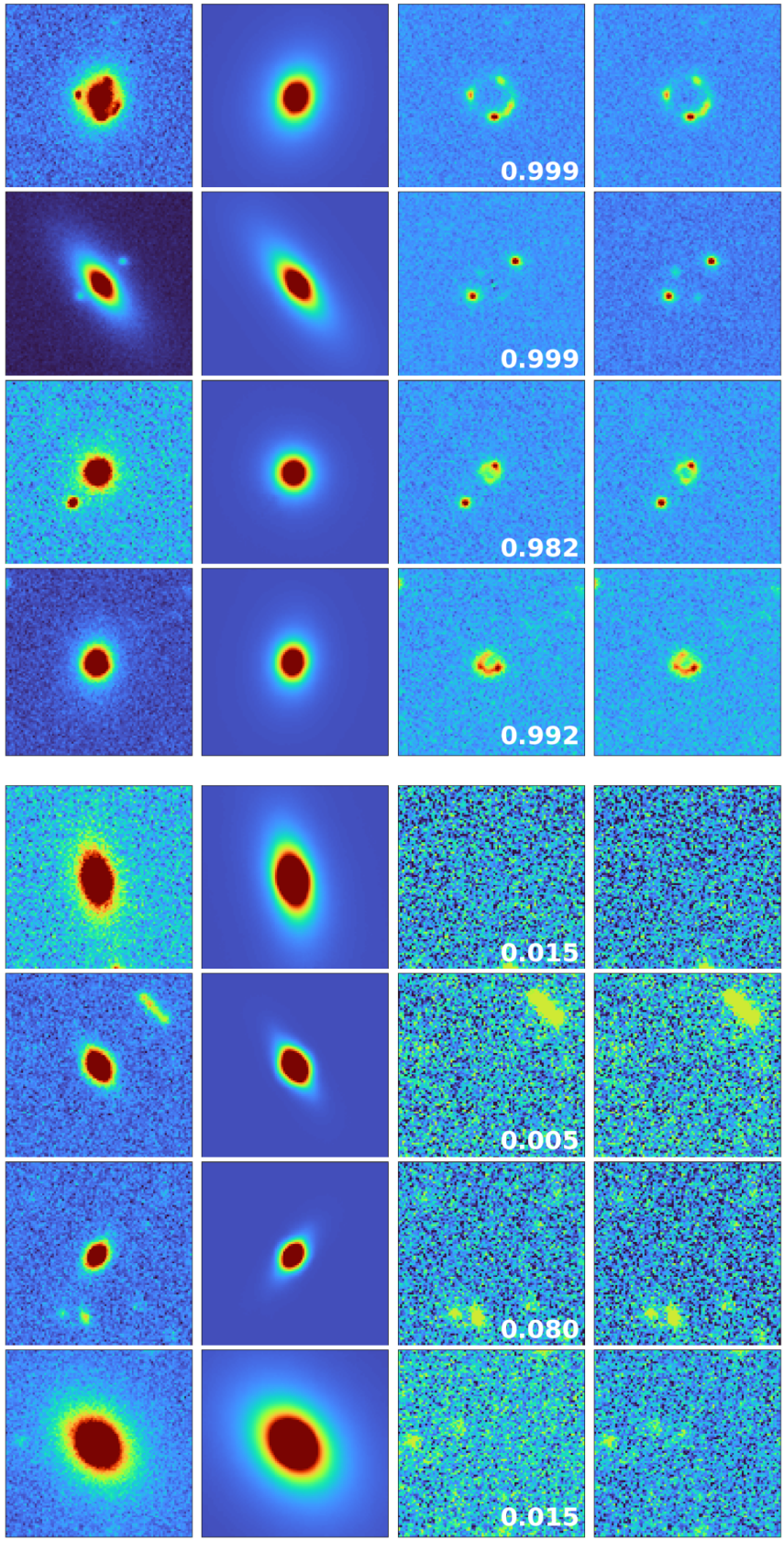}
    \caption{U-Net testing results on \texttt{Test\_sim}. Each row represents a set of prediction results, with the top four rows representing lenses and the bottom four rows representing non-lenses. First column: $r$-band image for the whole lens system or non-lens system. Second column: U-Net output. Third column: Residual of the previous two images. The white numbers signify the $P_{\textrm{ResNet}}$ (Sect. \ref{sec:resnet}). Fourth column: Simulated arc from the full lens system or background noise in non-lenses cases.}
    \label{fig:u_pred_simu}
\end{figure}

ResNet is a deep learning architecture that won the ImageNet Large Scale Visual Recognition Challenge (ILSVRC) 2015 competition \citealt{2016cvpr.confE...1H}. Its key innovation is the "skip connection," which enables the network to learn residual functions. This mechanism effectively overcomes the vanishing gradient problem, enabling training of extremely deep networks (hundreds of layers) while improving training speed and generalization. Due to its strong feature extraction and classification capabilities, ResNet has become a popular and effective tool in gravitational lens searches (e.g.,  \citealt{2019MNRAS.482..807P},  \citealt{2019ApJS..243...17J},  \citealt{2020ApJ...899...30L, 2021ApJ...923...16L} and others). In our methodology, the ResNet model serves as a binary classifier specifically designed to categorize the simpler residual images generated by the U-Net, rather than analyzing the original complex galaxy images. The training dataset consists of these residuals, where "positive" examples are derived from lensed images containing arc signals, and "negative" examples are from non-lensed images, which contain only noise. The network takes a single-channel ($r$-band) residual image as input. After training, the ResNet model makes final predictions by receiving a residual image from the test set and outputting a probability score ($P_{\textrm{ResNet}}$) between 0 and 1. This score indicates the model's confidence that the residual contains a genuine lensed arc signal; values closer to 1 indicate a higher probability that the original image represents a gravitational lensing system, thus marking it as a strong candidate for further investigation.

\subsection{Training the networks}
\label{subsec:training}
Before being input into the network, all training images (both inputs and outputs) underwent preprocessing and data augmentation. For U-Net training, images were cropped from $151 \times 151$ to $96 \times 96$, whereas the ResNet training images were further cropped to $85 \times 85$. This cropping procedure was designed to eliminate extraneous information at the edges of the images. We also applied random rotations of $0^{\circ}$, $90^{\circ}$, $180^{\circ}$, and $270^{\circ}$ to enhance the diversity of the training data. Additionally, to augment the ResNet dataset, we performed random shifts of less than 5 pixels in four directions. The pixel values were first squared and then normalized linearly to the range of 0 to 255. These preprocessing steps helped artificially expand the dataset, preventing model overfitting and improving generalization capability (\citealt{10.1145/3065386, 2020ApJ...899...30L, 2021ApJ...923...16L} and others). The U-Net was trained using the Adam optimizer and the Huber loss function, which is more robust against outliers than the mean squared error loss (\citealt{2015arXiv150408083G}). This choice is particularly advantageous for handling varying noise and features in astronomical images. The ResNet classifier was also trained with the Adam optimizer, utilizing a binary cross-entropy loss function, which is standard for binary classification tasks.

Initially, we trained the U-Net on the dataset \texttt{Train\_unet}, reserving 20\% of the data for validation. After each training epoch, the model's performance was evaluated on this nonoverlapping validation set to monitor learning progress. After completing the training, we applied the trained U-Net model to the dataset \texttt{Train\_res} to remove foreground galaxies from all images. The resulting residuals were then used to train the ResNet model. Finally, both the U-Net and ResNet were tested using the datasets \texttt{Test\_sim} and \texttt{Test\_real} to assess their performance.

\begin{figure*}
    \centering
    \includegraphics[width=0.32 \textwidth]{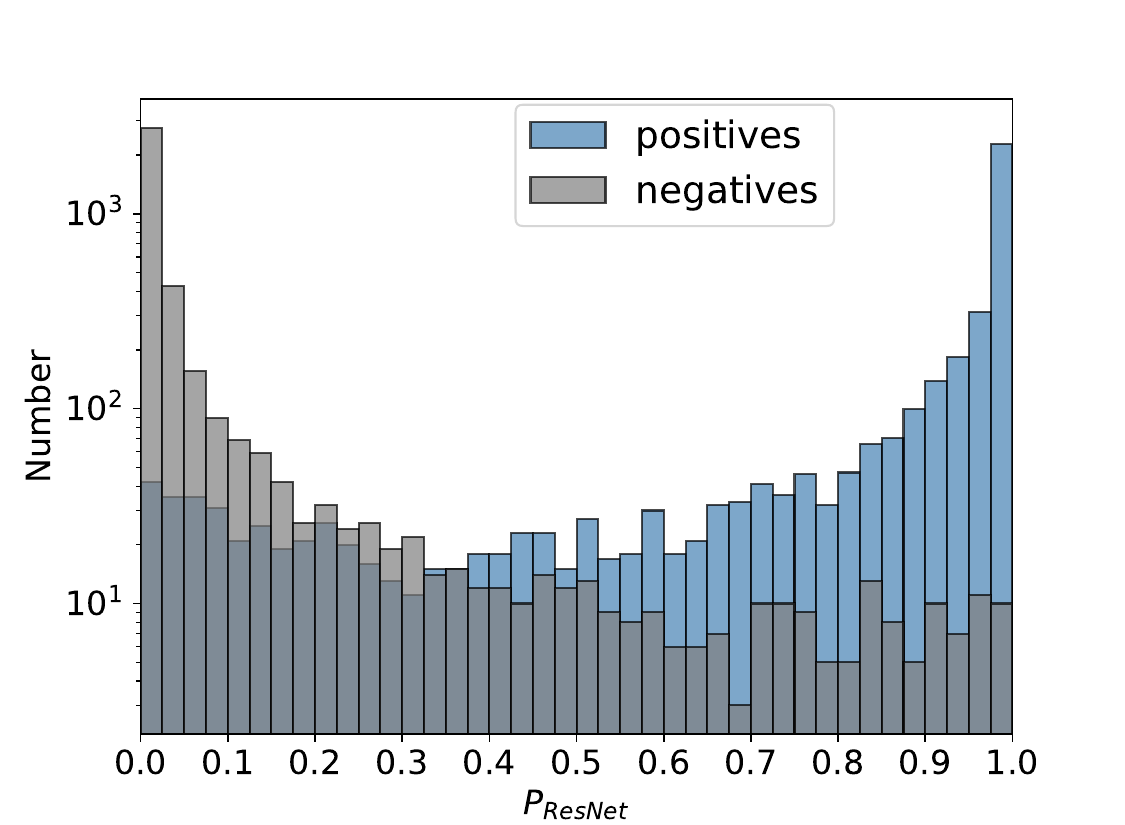}
    \includegraphics[width=0.32 \textwidth]{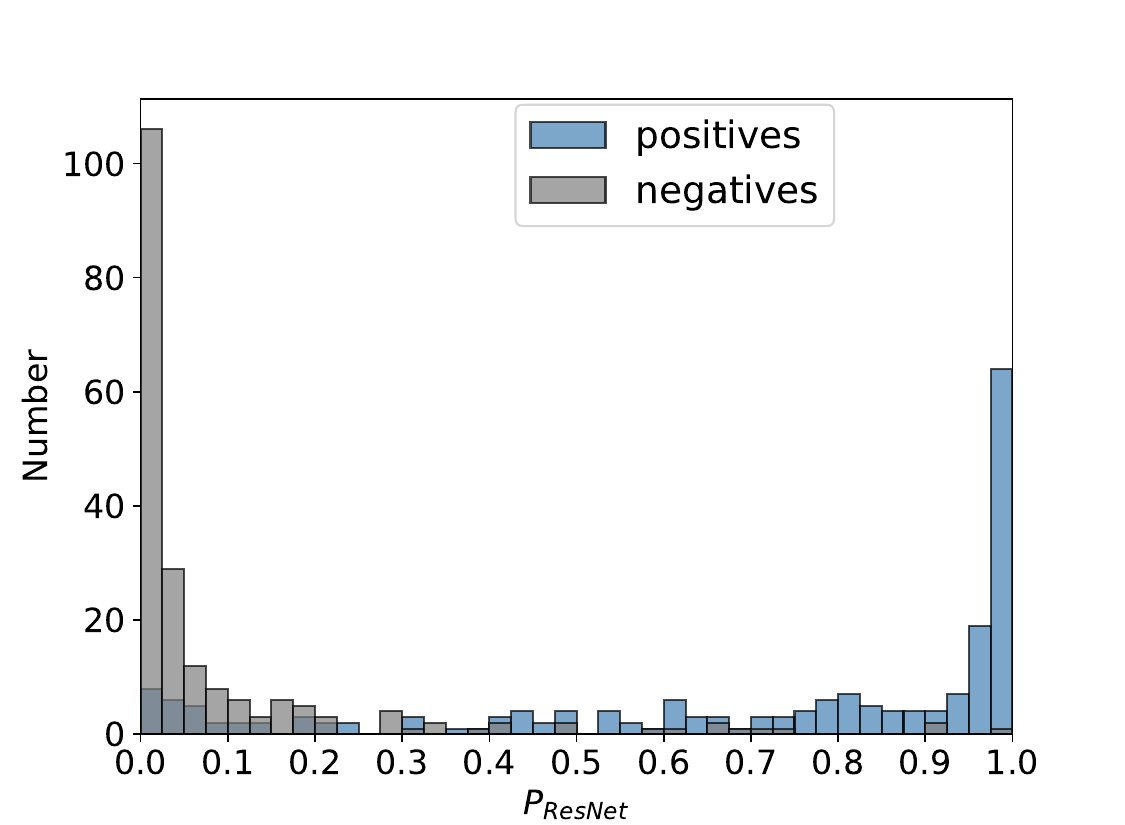}
    \includegraphics[width=0.32 \textwidth]{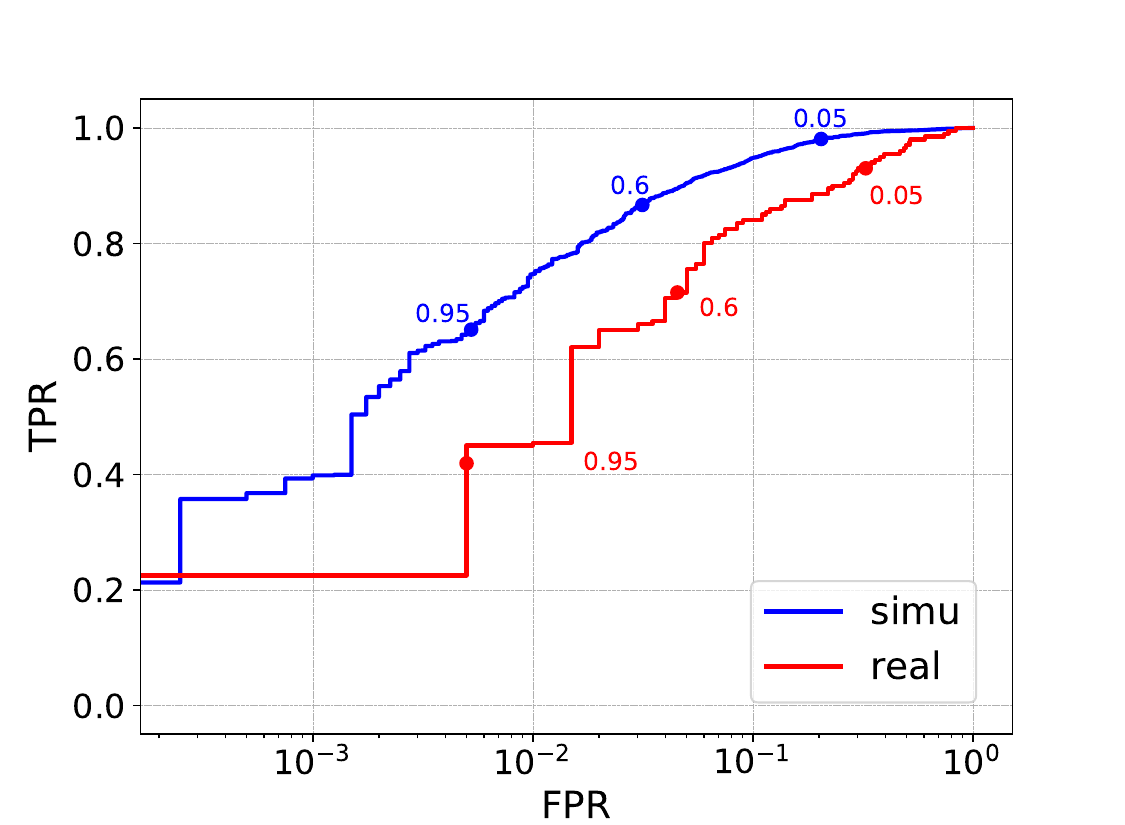}
    \caption{ResNet testing results. Left panel: Probability distribution on \texttt{Test\_sim}. Middle panel: Probability distribution on \texttt{Test\_real}. Right panel: ROC curve of \texttt{Test\_sim} and \texttt{Test\_real} with several probability threshold points labeled.}
    \label{fig:testresult}
\end{figure*}

\begin{figure}[ht!]
\centering
\includegraphics[width=\hsize]{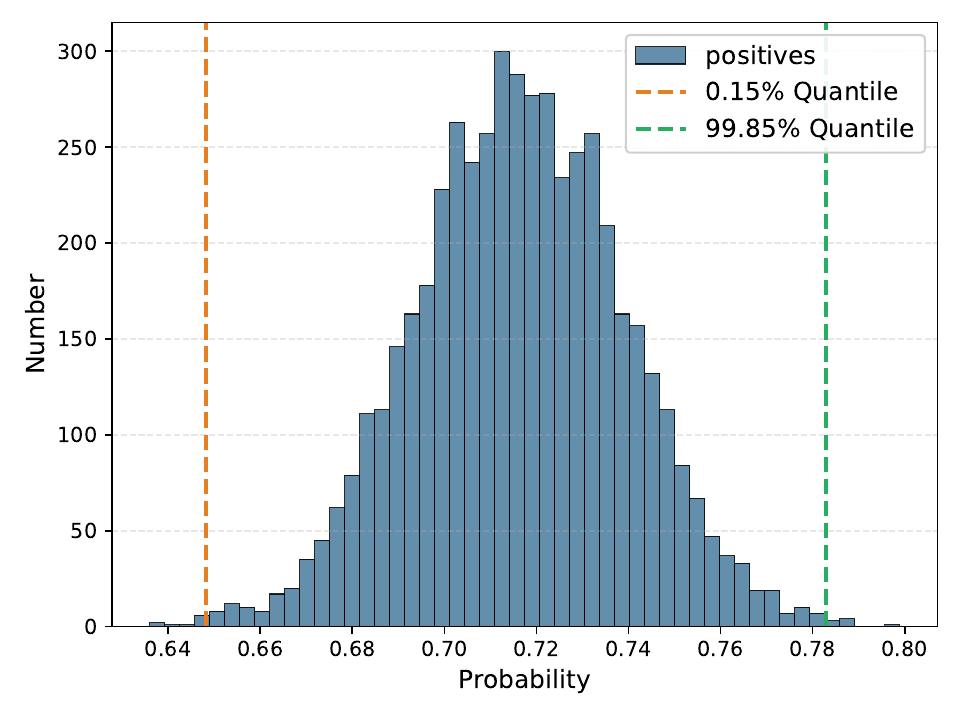}
    \caption{Bootstrapping test results on \texttt{Test\_real} positive data. We randomly selected 5,000 groups with replacement and set the statistical range to $0.15\%-99.85\%$.}
    \label{fig:bootstrapping}
\end{figure}

\section{Performance of the two networks}
\label{sec:performance}
In this section, we validate the performance of the U-Net and ResNet, as well as the effectiveness of the U-Net to ResNet framework. We conduct the evaluation using two nonoverlapping datasets, ensuring the integrity of our analyses. As described in Sect. \ref{sec:data_preparation}, the first dataset, \texttt{Test\_sim}, consists of simulated lenses that inject simulated arcs into cutouts containing real central galaxies. This dataset is critical for assessing the intrinsic capabilities of our model under controlled conditions. The second dataset, \texttt{Test\_real}, comprises real KiDS cutouts, with positive candidates primarily derived from known high-quality lens candidates previously identified in the KiDS dataset. These candidates are paired with non-lensing examples.

To qualitatively analyze our results, we define a set of metrics fundamental to assessing model performance. The standard counts of true positives (TPs), false positives (FPs), true negatives (TNs), and false negatives (FNs) at a specified probability threshold \(p\) provide a foundation for calculating key performance indicators. The true positive rate (TPR), also known as recall or sensitivity, measures the proportion of actual positives correctly identified by the model and is defined as 
\[
    \text{TPR} = \frac{\text{TP}}{\text{TP} + \text{FN}}.
\]
A higher TPR indicates that the model effectively identifies positive cases. The false positive rate (FPR), often referred to as contamination, quantifies the proportion of actual negatives that are incorrectly classified as positives, and is defined as 
\[
    \text{FPR} = \frac{\text{FP}}{\text{FP} + \text{TN}}.
\]
A lower FPR indicates better model performance in distinguishing between positives and negatives. Furthermore, we evaluated the receiver operating characteristic (ROC) curve, which illustrates the trade-off between the TPR and FPR across varying thresholds \(p\). The ROC curve plots the TPR against the FPR for different probability thresholds, enabling us to visualize the model's performance across different sensitivity levels. In some instances, we also report the absolute number of machine candidates above a designated threshold, as the costs associated with human vetting correlate with the number of candidates generated. It is important to note that most samples in \texttt{Test\_real} are lens candidates, and the exact number of confirmed gravitational lenses remains unknown. Direct assessments of the TPR and FPR are strictly applicable only to simulations. For the tests with real data, we analyzed the performance on the candidates to provide a meaningful evaluation of our model's effectiveness; however, we cannot guarantee that the resulting metrics reflect actual accuracy, as the samples used consist solely of candidates.

A critical function of the U-Net is to remove central light while preserving the residuals characteristic of various strong lensing phenomena. Utilizing the simulation dataset \texttt{Test\_sim}, we demonstrate that the U-Net effectively eliminates central light from both lensing and normal galaxies. In Fig. \ref{fig:u_pred_simu}, we present several examples of strong lenses, including Einstein rings, Einstein crosses, normal arcs, and arcs within complex environments such as galaxy clusters. For each example, we show the raw image, the extracted foreground light, and the residual obtained by subtracting the predicted foreground light from the raw image. Additionally, we include simulated arcs that have been directly injected into the background noise for comparison. Our results indicate that the U-Net produces high-quality residuals for all types of lenses. When comparing these residuals to the simulations, we do not observe any significant differences, suggesting that our model performs reliably in these scenarios. This finding extends to the removal of foreground light from normal galaxies, as seen in the last four rows of Fig. \ref{fig:u_pred_simu}. In these cases, the residuals appear very similar to the actual background noise, indicating the model's capability to accurately manage light from various sources.

When it comes to real observed data, the U-Net performs well. In the left panel of Fig. \ref{fig:u_pred_real} and Fig. \ref{fig:u_pred_real_apd}, we present cases of real lenses and galaxies selected from the dataset \texttt{Test\_real}. For the strong lenses depicted in the first three cases (Fig. \ref{fig:u_pred_real}), the environments are clear, and the U-Net effectively removes the foreground light, yielding distinctly visible lensed images. This clarity is evident not only in the regions surrounding the lenses but also in the centers where the foreground galaxies are located; in these areas, we observe only noise, indicating that all signals from the central galaxies have been successfully removed. In the next three examples (top three rows in Fig. \ref{fig:u_pred_real_apd}), we showcase successful cases set within complex environments. Despite the crowded surroundings, the U-Net successfully subtracts the foreground light, yielding satisfactory residuals for the lensed images. However, in the last two cases (bottom two rows in Fig. \ref{fig:u_pred_real_apd}), we present instances of failure. Some of these cases involve multiple central galaxies. In such scenarios, the U-Net may misinterpret one of the central galaxies as a lensed image, which can lead to an incorrect prediction of the foreground light profile. Consequently, this results in erroneous lensed arcs that do not conform to expected gravitational lensing patterns. Furthermore, some of these cases may not involve genuine lenses at all. In the residuals of such systems, we do not observe lensed images that align with typical lens characteristics, suggesting that the model's assumptions may not have been met. In the right panel of Fig. \ref{fig:u_pred_real} and Fig. \ref{fig:u_pred_real_apd}, we demonstrate the model's performance on real galaxies. Similar to the lensing cases, the first three rows (Fig. \ref{fig:u_pred_real}) show galaxies in clear environments, while the next three cases illustrate galaxies situated in crowded conditions where reconstruction remains successful (top three rows in Fig. \ref{fig:u_pred_real_apd}). However, the last two cases also represent failures (bottom two rows in Fig. \ref{fig:u_pred_real_apd}). In these instances, the central galaxies may be disk galaxies or exhibit substructures, such as spiral arms or bulges similar to arc structures. These complexities are challenging for the U-Net, as a single S\'ersic profile cannot adequately capture the diverse light distributions present in such galaxies. For these types of complexities, we must acknowledge the current limitations of our model, as our present simulation methods are unable to effectively model such intricately structured galaxies.

Next, we examine the performance of the ResNet. As shown in the left and middle panels of Fig. \ref{fig:testresult}, these illustrate the distribution of predicted probabilities generated by ResNet on the dataset \texttt{Test\_sim\_res} and \texttt{Test\_real}. This distribution provides an overall assessment of the model's predictive ability. In an ideal scenario, all positive cases should cluster around $P_{\textrm{ResNet}}=1$, while all negative cases should peak at $P_{\textrm{ResNet}}=0$. The closer the CNN's classification of these two categories is to this ideal scenario, the better its performance. The left panel of Fig. \ref{fig:testresult} shows the probability distribution for simulation data, \texttt{Test\_sim\_res}. We observe that the majority of lenses are positioned on the right, while most normal galaxies are found on the left. The CNN successfully differentiates between these two populations, as indicated by the scarcity of samples in the intermediate probability range (e.g., 0.2–0.8). This means that the model is quite confident in distinguishing between the two classes. When we move to real data, the performance is slightly worse; however, the majority of the two groups still concentrate on their respective sides. In comparison to the simulations, we notice that for the positive cases, the distribution appears somewhat broader from right to left. This suggests that some lenses may be missed in the real data compared to the simulations. It is also important to consider alternative explanations. Given that the lenses in the real data are actually only candidates, they may be FPs. The new model can correctly classify these candidates, even if previous works considered them as genuine lenses.

Turning to the right panel of Fig. \ref{fig:testresult}, we display the TPR against the FPR for the ResNet model. Given the small sample size of \texttt{Test\_real}, we employed a bootstrapping method to determine a robust probability segmentation threshold. We performed 5,000 resampling iterations. In each iteration, 200 probabilities of the positive sample prediction were randomly drawn with replacement from the \texttt{Test\_real} dataset, and their mean was calculated. The distribution of these 5,000 calculated means, plotted in Fig. \ref{fig:bootstrapping}, can be approximated by a normal distribution with a mean $\mu = 0.715$ and a standard deviation $\sigma = 0.022$. The orange and green lines indicate the boundaries of the 99.7\% confidence interval, corresponding to $\mu \pm 3\sigma$. We selected a segmentation threshold of 0.6. This value is below $\mu - 3\sigma$ and encompasses the full range of the bootstrapped probability distribution. At this threshold, on \texttt{Test\_real}, our model achieves an FPR of 4.5\% alongside a TPR (completeness) of 71.5\%. While these metrics may initially appear modest, particularly when compared to the well-documented performance of earlier KiDS studies, it is crucial to recognize the selection bias inherent in \texttt{Test\_real}. Our model was specifically trained to identify arcs that are hidden by foreground galaxies, which are usually small, faint, and often possess modest S/N. In contrast, many of the confirmed, higher-grade candidates present within \texttt{Test\_real} originate from pipelines optimized for brighter, wider-separation lensing features. For instance, the initial KiDS CNN presented by  \citealt{2017MNRAS.472.1129P} targeted $R_{\textrm{E}}$ values greater than $1.4''$, effectively favoring wide-separation and high-contrast lenses that are more easily detectable by standard detection methodologies. Subsequent studies, including those by  \citealt{2019MNRAS.484.3879P} and  \citealt{2021ApJ...923...16L}, improved their detection efficacy by transitioning from single-band to three-band convolutional networks and enhancing their training data and strategies, while still focusing on visually prominent systems. This discrepancy elucidates the observed lower TPR of our model on \texttt{Test\_real}, despite its robust performance on the earlier, controlled \texttt{Test\_sim} dataset.

Our tests suggest that while the U-Net to ResNet pipeline demonstrates strong capabilities for our target lenses, there are still opportunities to refine the model’s applicability to brighter lens classes. This can be achieved through training on additional simulated data or introducing multiband color channels. These improvements could strengthen our approach in the rapidly evolving landscape of machine learning applications for astrophysical lens detection.

\begin{figure}[ht!]
\centering
\includegraphics[width=\hsize]{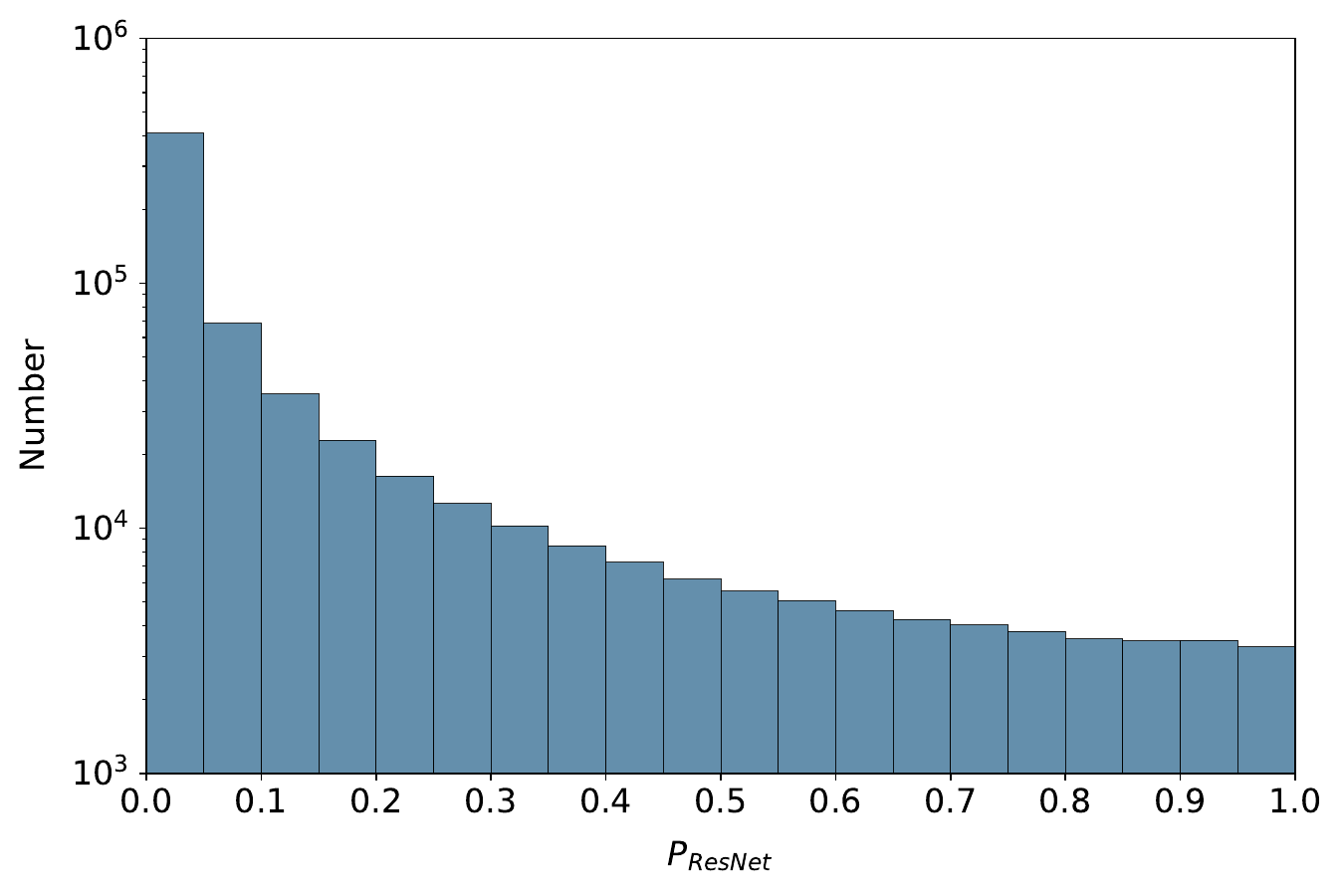}
    \caption{Predictive data probability distribution. This figure shows the probability distribution on predictive data. The total number of the predictive data points is 638,398.}
    \label{fig:application}
\end{figure}

\begin{figure*}
        \centering
        \includegraphics[width=0.23 \textwidth]{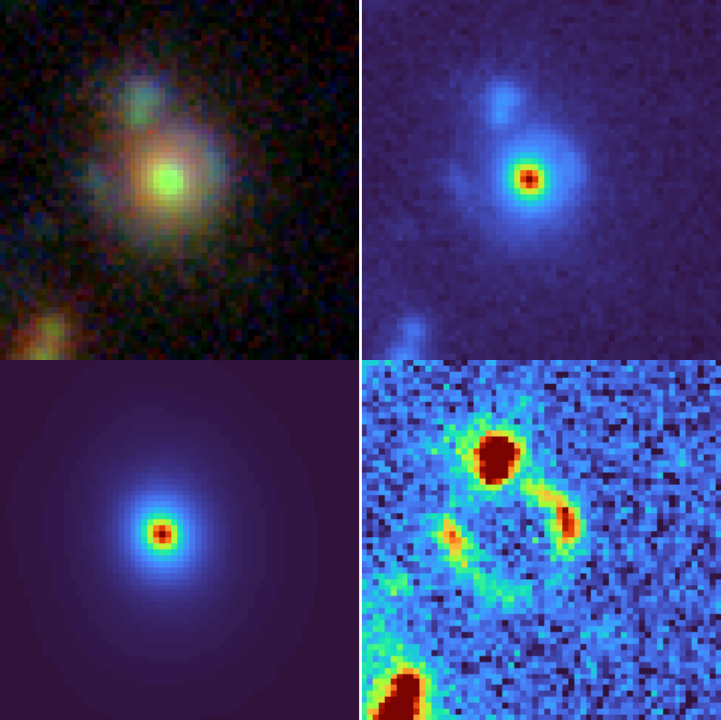}
        \quad
        \includegraphics[width=0.23 \textwidth]{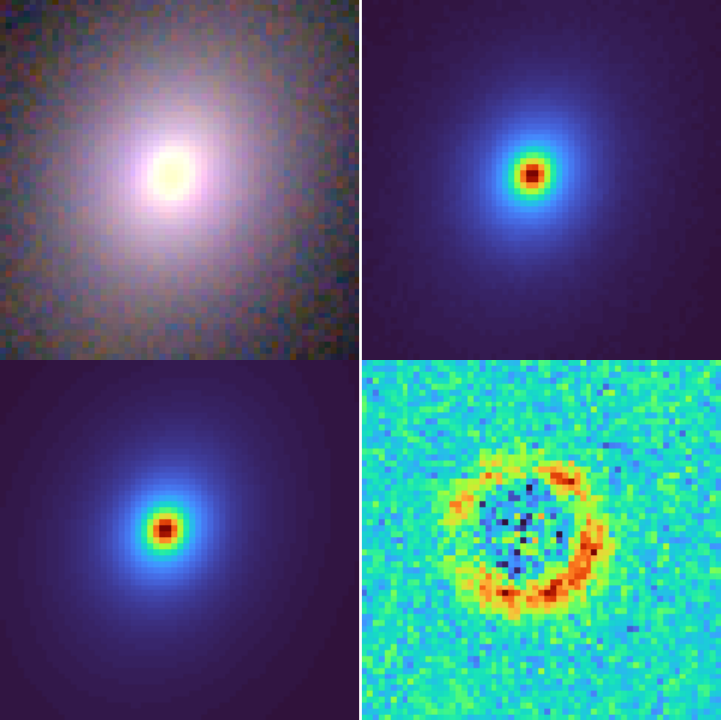}
        \quad
        \includegraphics[width=0.23 \textwidth]{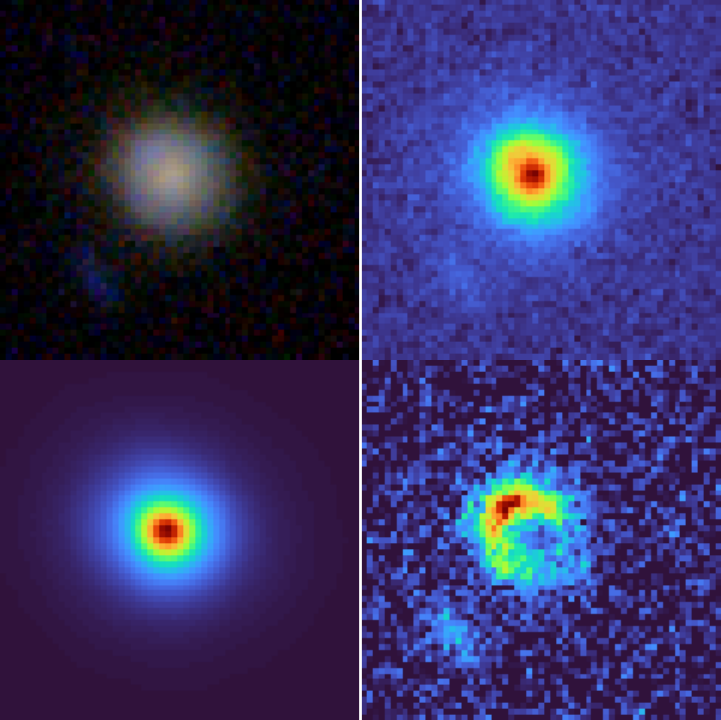}
        \quad
        \includegraphics[width=0.23 \textwidth]{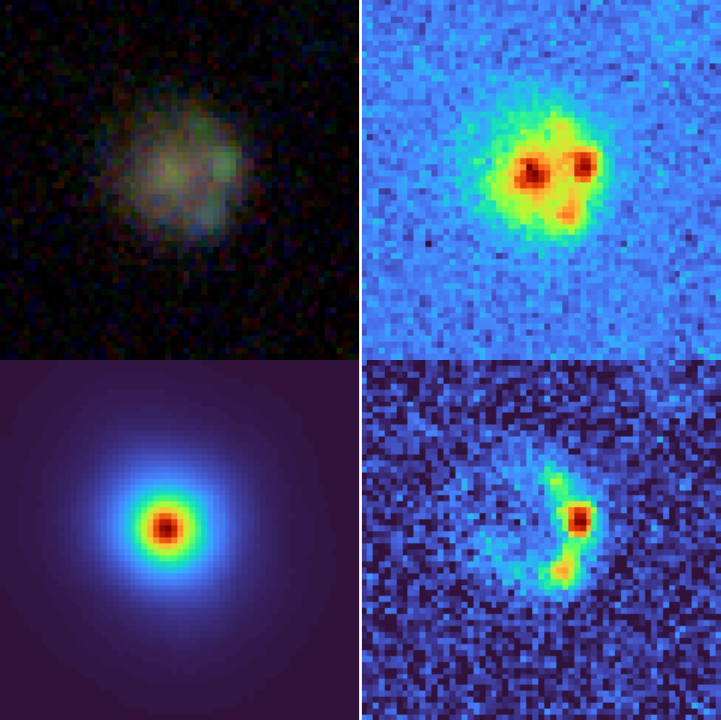}
        \newline

        \includegraphics[width=0.23 \textwidth]{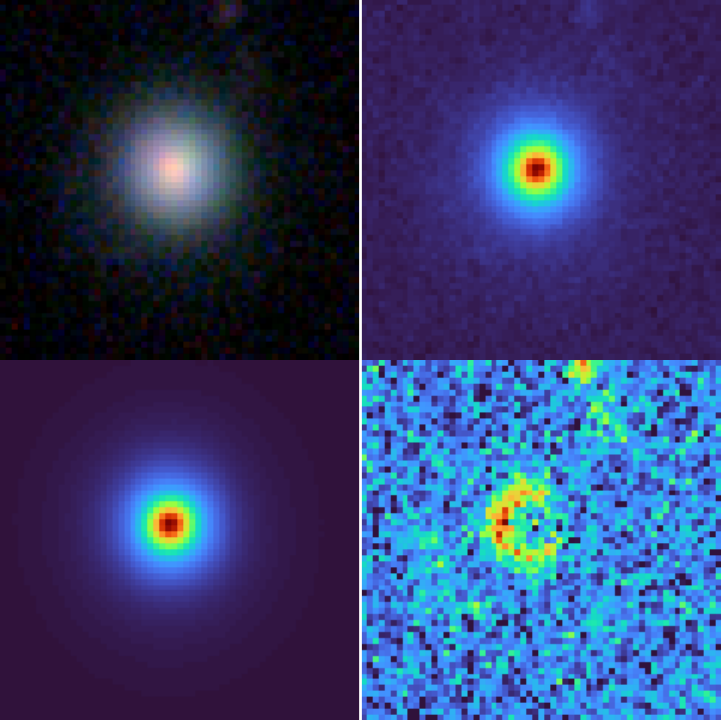}
        \quad
        \includegraphics[width=0.23 \textwidth]{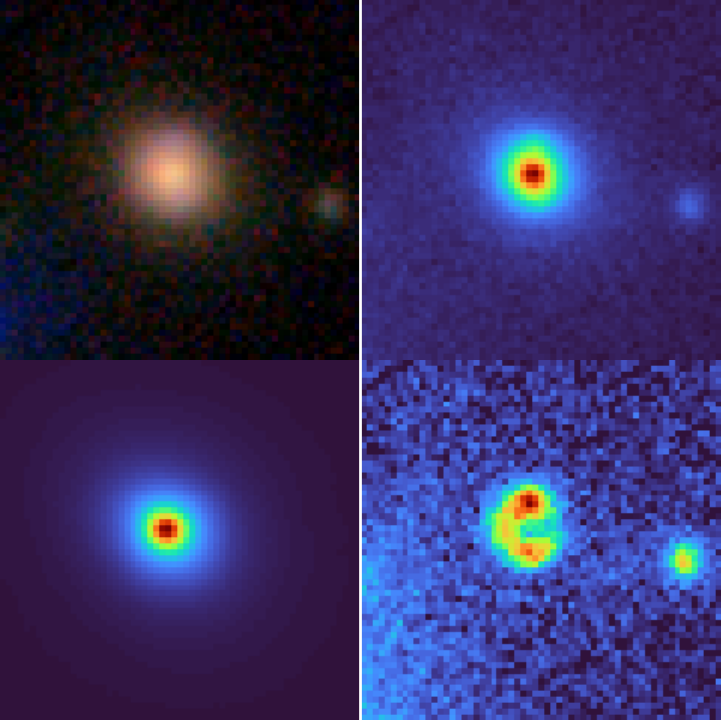}
        \quad
        \includegraphics[width=0.23 \textwidth]{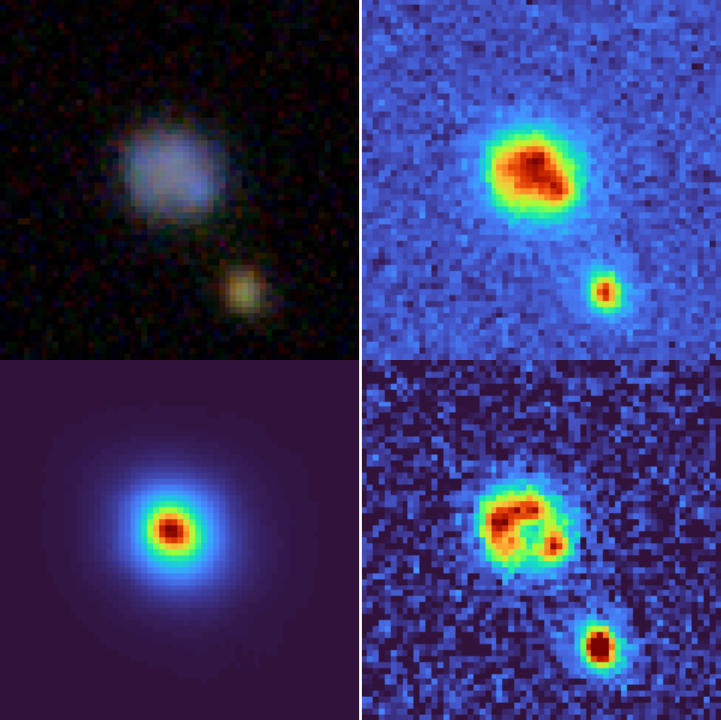}
        \quad
        \includegraphics[width=0.23 \textwidth]{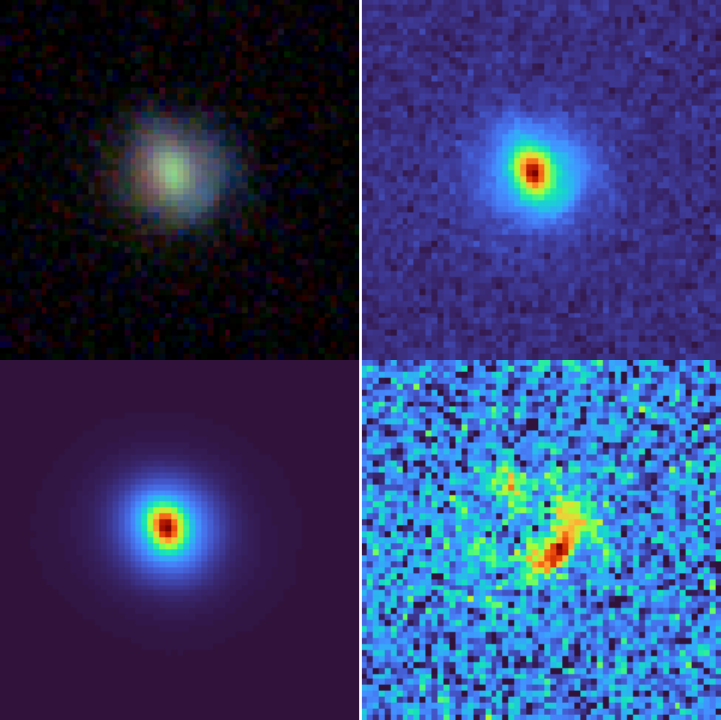}
        \newline

        \includegraphics[width=0.23 \textwidth]{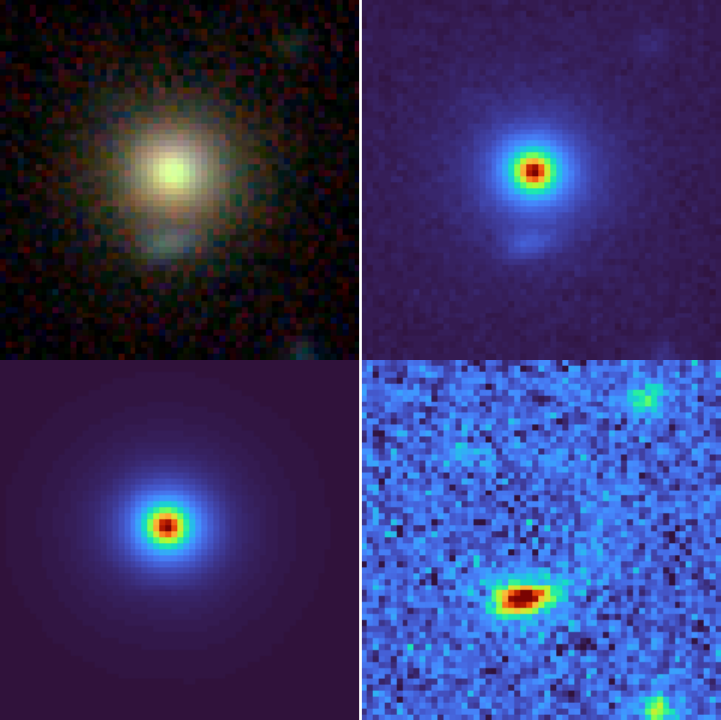}
        \quad
        \includegraphics[width=0.23 \textwidth]{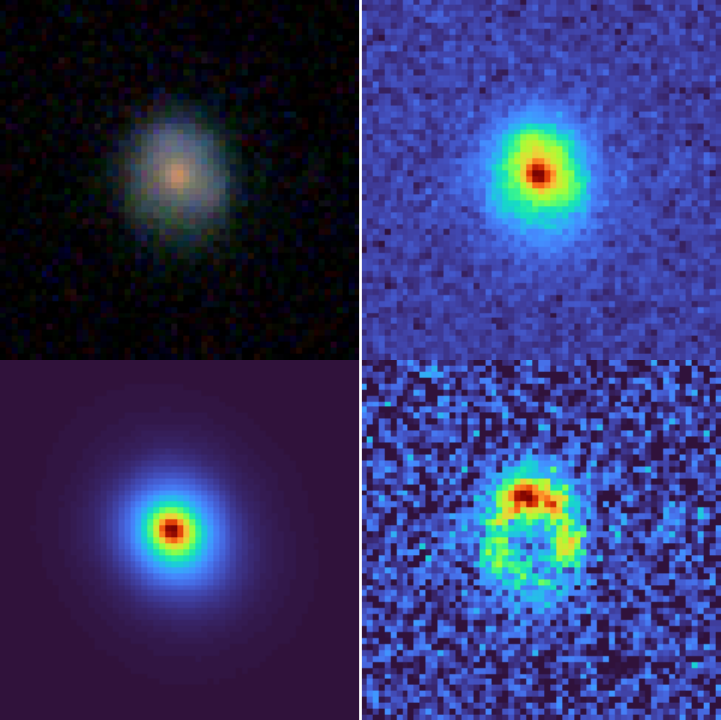}
        \quad
        \includegraphics[width=0.23 \textwidth]{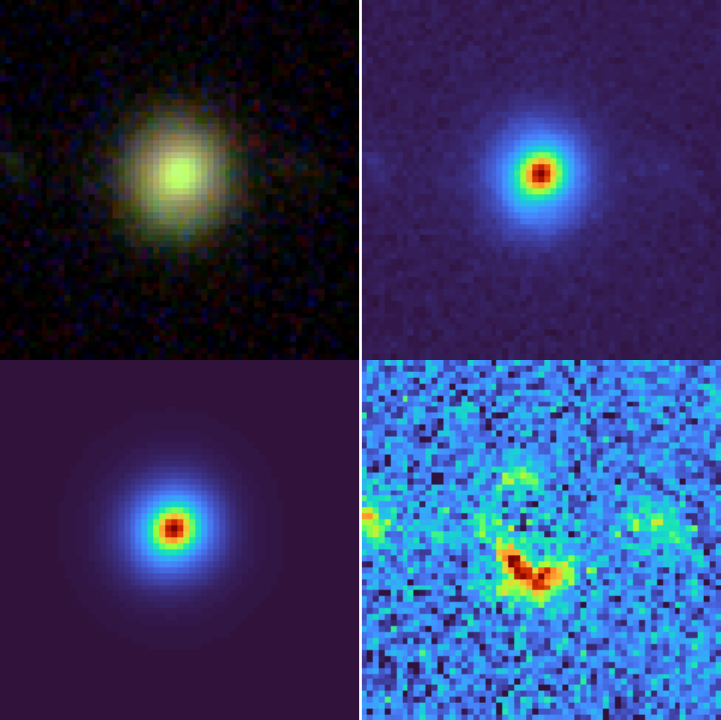}
        \quad
        \includegraphics[width=0.23 \textwidth]{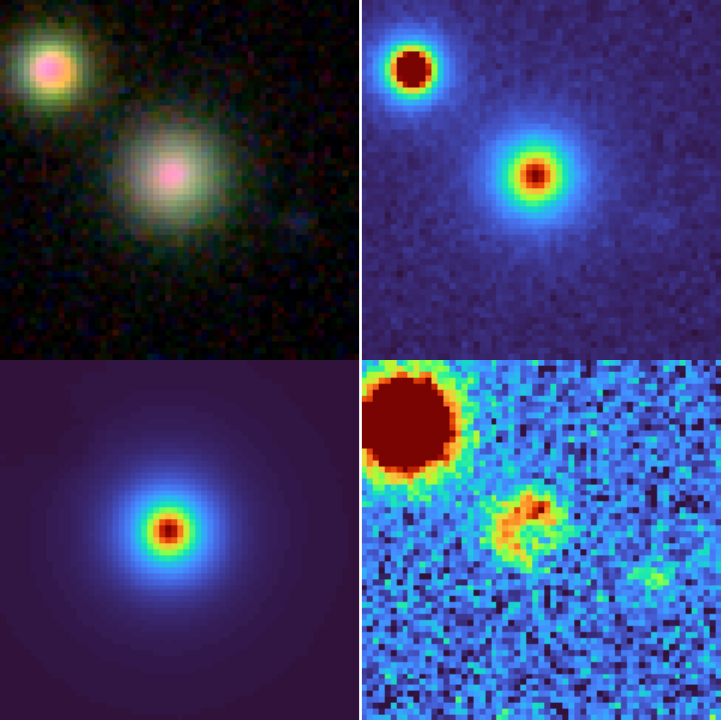}
        \newline

        \includegraphics[width=0.23 \textwidth]{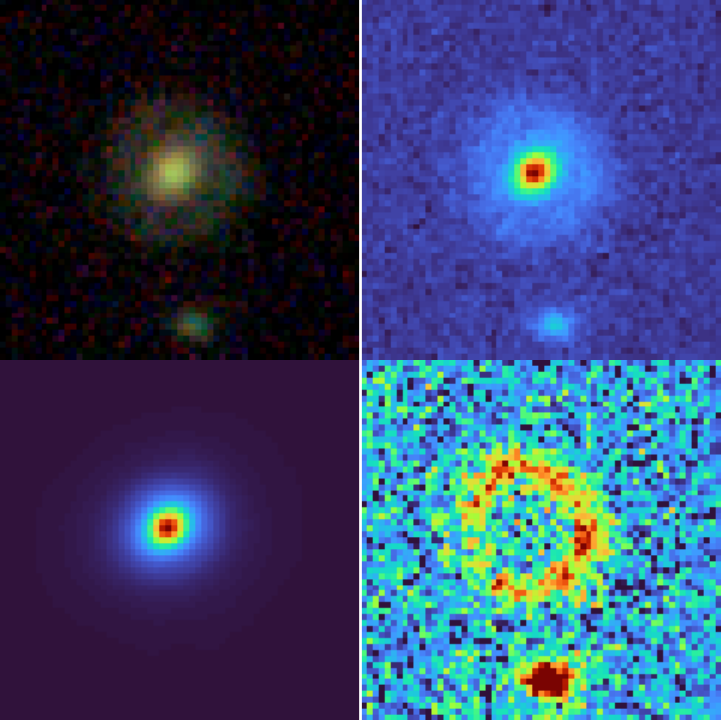}
        \quad
        \includegraphics[width=0.23 \textwidth]{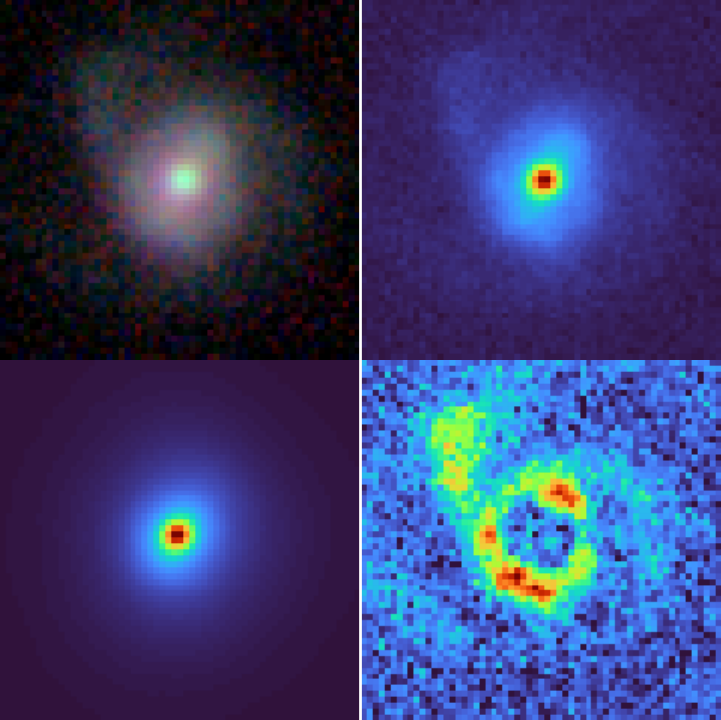}
        \quad
        \includegraphics[width=0.23 \textwidth]{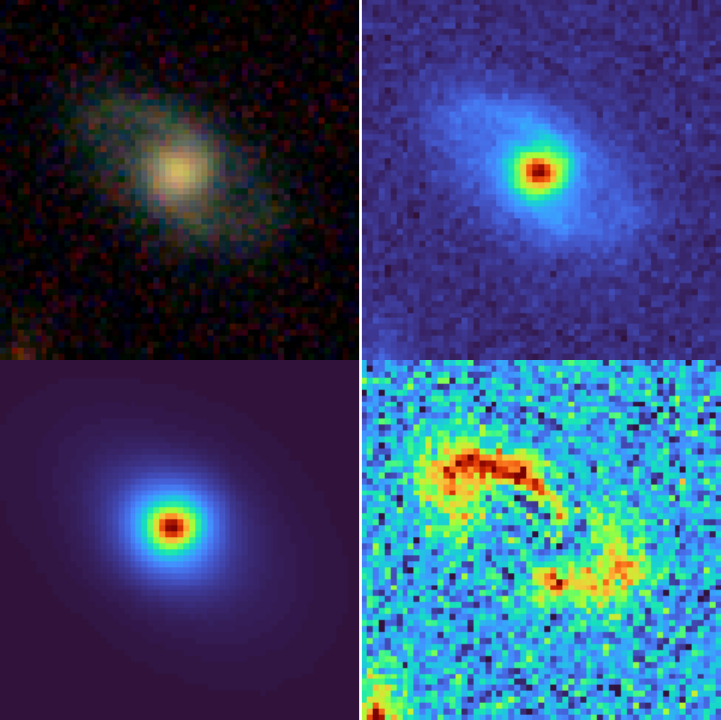}
        \quad
        \includegraphics[width=0.23 \textwidth]{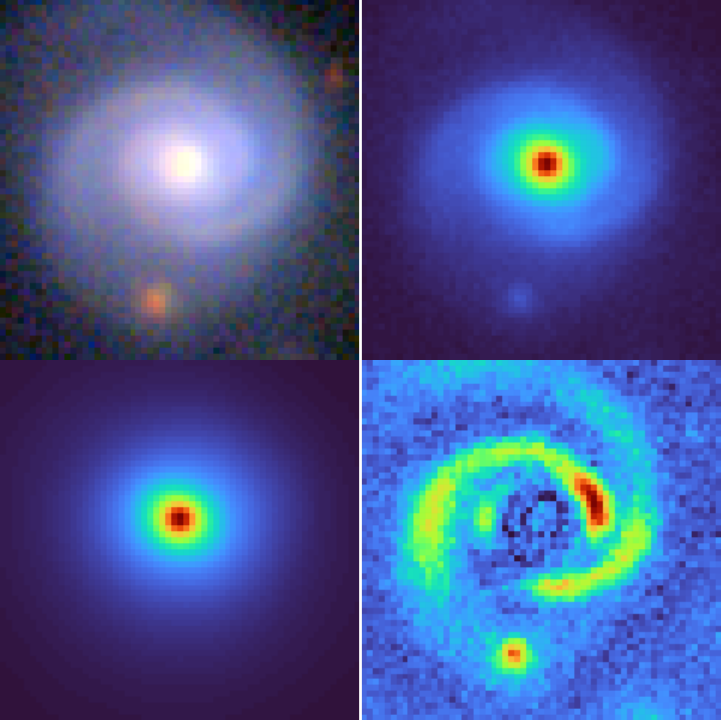}

        \caption{Representative samples of high-probability candidates ($P_{\mathrm{ResNet}} > 0.6$) separated by their human classification. 
Top to bottom: Class A, class B, class C, and discarded FPs. 
Each example is presented as a four-panel, $2\times2$ mosaic layout. 
(1) Upper-left: $g$, $r$, and $i$ color-composite image of the system.
(2) Upper-right: Original KiDS $r$-band observation.
(3) Lower-left: Foreground galaxy model predicted by U-Net (in $r$-band).
(4) Lower-right: Final residual image after subtracting the U-Net model from the original $r$-band data.
Note that with the exception of the color-composite image (upper-left), all other panels display data derived exclusively from the $r$-band.}
        \label{fig:candidates}
    \end{figure*}

\begin{figure*}[htbp!]
    \centering
    \vspace{3mm}
    \includegraphics[width=1 \textwidth]{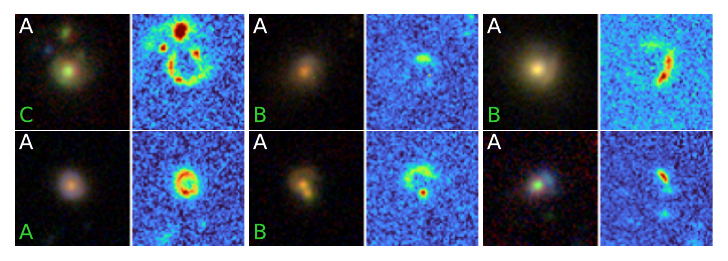}
    \caption{Subset of candidates successfully matched with the Master Lens Database. The remaining candidates are shown in Fig. ~\ref{fig:match_apd}. Each subplot displays the color image on the left and the corresponding $r$-band residual image on the right. 
    The white text in the upper-left corner indicates the classification grade assigned in this work. When present, the green text in the lower-left corner gives the visual scoring grade from the previous study, as described in Sect.~\ref{subsec:visual_classification}.}
    \label{fig:match}
\end{figure*}

\section{Application to KiDS-DR4 data}
\label{sec:real-application}
After testing the U-Net subtraction and ResNet classification pipeline on simulations and real data, we applied this methodology to observational data from KIDS-DR4. Transitioning from simulations to real-world survey data presents significant challenges, mainly due to the complexity of foreground morphologies and the presence of non-lensing contaminants. The compilation of our final strong lens catalog follows a systematic workflow. This involves defining a scientifically relevant predictive sample, executing the deep learning inference and conducting a multistage human visual inspection to refine the candidate selection.

\subsection{Data selection and model inference}
\label{subsec:data_inference}
The KiDS-DR4 provides multiband photometry in nine filters; however, we chose the $r$-band images for our main analysis due to their superior image quality, characterized by a median full width at half maximum (FWHM) of approximately $0.7''$. This high resolution is critical for detecting lensed features that blend with the light from foreground galaxies, particularly those with small $R_{\mathrm{E}}$. The sharpness of the $r$-band images helps reveal features that might be overlooked in lower-resolution bands. 

In our study of gravitational lensing, it is crucial to recognize that we cannot apply our models to every object in our survey. Specifically, low-redshift dwarf galaxies, stars, and quasars lack the mass necessary to produce strong gravitational lensing effects. Therefore, we selectively chose objects more likely to serve as effective gravitational lenses, enabling us to focus our efforts and improve the accuracy of our results. First, we removed the targets classified as stars and quasars, as identified by  \citealt{2025ApJS..279...26F}, and retained only those with a galaxy probability (\(p_{gal}\)) greater than 0.5. This approach reduces the model's running time and effectively decreases the FPs associated with stars and quasars.  
Second, we imposed a magnitude cutoff, requiring \(r\)-band magnitudes from the SExtractor catalog (\citealt{1996A&AS..117..393B}) to be brighter than 20 (\(r_{\mathrm{mag\_auto}} < 20\)). Although this criterion might exclude some faint lensing candidates, it ensures a sufficiently high S/N for the foreground light. This high S/N is crucial for our U-Net model to accurately separate light from foreground objects. Next, we applied a stellar mass criterion, selecting galaxies with masses greater than \(10^{10.5} \text{M}_\odot\), as lighter galaxies are unlikely to form detectable strong gravitational lenses. This mass limit is based on stellar mass estimates obtained with \texttt{LePhare} using the method described in \citealt{2022irbf.confE..16X} by assuming the Chabrier initial mass function. These sequential filtering steps significantly reduced the size of our predictive dataset. From an initial sample of 13,561,935 galaxies, the application of both magnitude and mass selection criteria yields a final sample of 638,398 massive galaxies. We acknowledge that this selection entails a trade-off in completeness. However, it substantially reduces the FPR and significantly improves the efficiency of visual inspection. This sample balances broad coverage with a focused selection, targeting objects most likely to act as gravitational lenses. We cross-matched this predictive sample with known lenses in the KiDS. From 366 candidate lenses identified in various previous studies (\citealt{2019MNRAS.484.3879P, 2020ApJ...899...30L, 2021ApJ...923...16L}), we complied a catalog of 275 known strong lenses present in our sample.

These 638,398 targets were then input into the U-Net, from which the foreground galaxy light is removed. The ResNet model assigns a probability score, \(P_{\mathrm{ResNet}}\), to the residuals to indicate lensing features. According to the tests described in Sect. \ref{sec:performance}, we find that \(P_{\mathrm{ResNet}} > 0.6\) yields a completeness of over 71.5\%, while maintaining a low FPR of 4.5\% in our tests. This threshold was established by balancing the TPR and FPR, informed by metrics such as the ROC curves from our training data. Applying this threshold to the full sample yields 30,456 candidates (i.e., the model predicting result in Fig. \ref{fig:application}), ready for further evaluation. For the 275 known strong lenses (mentioned in Sect. \ref{subsec:data_inference}), the pipeline successfully identifies 171 of them above the 0.6 threshold, resulting in a recovery rate of approximately 62\% (\(171/275\)). This rate serves as a useful benchmark, indicating that our method aligns well with human inspections and supports its application in large surveys. However, we consider this 62\% recovery rate as a lower limit, because, as discussed above, known lens catalogs are biased toward systems with larger $R_{\mathrm{E}}$, which are easier to detect without subtraction due to their broader arcs.

\subsection{Candidate selection and visual classification}
\label{subsec:visual_classification}
The 30,456 machine-selected candidates, after removing the known ones, were subjected to a two-stage visual inspection to eliminate FPs. To facilitate this inspection, we generated a comprehensive \(2\times2\) diagnostic mosaic for each candidate (see Fig. \ref{fig:candidates}). This mosaic visualizes the system in the multiband color space (\(g, r, i\)), the original high-resolution \(r\)-band image, the U-Net reconstructed foreground model, and the foreground-subtracted residual map. The visual classification process involved a joint analysis of these panels to determine whether the candidates were strong lensing candidates. The initial stage consisted of a rapid visual inspection based on the foreground-subtracted residual images and the corresponding color-composite images. The residual images produced by the U-Net architecture effectively mitigate the impact of the foreground light. This capability substantially enhances the quality of visual inspection. Our goal in this stage was to remove objects that could be confidently identified as non-lenses based on their morphology. During this step, we discarded obvious contaminants, such as projected galaxy pairs, normal spiral galaxies with arms, ring galaxies, irregular galaxies, and other systems whose features were clearly unrelated to lensing. After this coarse screening, 2,168 candidates remained\footnote{The candidate catalog and the corresponding visual inspection mosaic are  available at \url{https://junjijia.whu.
edu.cn/lenses.zip}.}. We then conducted a second, rigorous visual inspection of the 2,168 candidates that survived the initial screening. We categorized these candidates into three classes (A, B, and C) based on the confidence level of the identification. The physical and morphological criteria for each class are detailed below:
\begin{description}
    \item[Class A:] 
    As shown in the top row of Fig. \ref{fig:candidates}, these candidates provide the strongest evidence for gravitational lensing. Morphologically, they display clear, high-surface-brightness arcs or multiple images in the residual maps, aligned tangentially to the center of the deflector. Notably, distinctive features are also evident in the color-composite images (upper-left panel). In these systems, the lensed sources often appear as blue, star-forming galaxies, providing a sharp contrast with the reddish central massive elliptical foreground galaxy. For Class A objects, the U-Net subtraction primarily serves as a validation tool, enhancing the contrast of the arcs and confirming that the geometry of the system is consistent with that of a lens. These candidates are classified as our "high-confidence sample" and represent the primary targets for spectroscopic confirmation to determine source redshifts.
    
    \item[Class B:] 
    As shown in the second row of Fig. \ref{fig:candidates}, these candidates represent the most significant scientific contribution of our pipeline. These systems are characterized by a complete lack of visible lensing features in the original observational data, where the faint light of the background source is entirely overwhelmed by the brightness of the central bulge of the foreground galaxy. However, the U-Net residual maps (lower-right panel) reveal compelling structures, such as counter-images or symmetric arcs. Physically, these candidates likely correspond to "small angle lenses" or systems with high flux ratios between the lens and the source. The identification of these candidates demonstrates our foreground-subtraction method's unique capability to probe a population of lenses inaccessible to traditional color- or magnitude-based searches.
    
    \item[Class C:] 
    The third row of Fig. \ref{fig:candidates} illustrates the low-probability cases. Unlike Class B, where the residual features contain multiple strong lensed images forming a geometrically coherent system (e.g., counter-images), Class C candidates typically display only a single plausible arc-like structure. While these features may represent lensed images of faint background sources, they could also stem from other physical phenomena. For instance, tidal debris, faint satellite galaxies, or even imperfect PSF modeling during the U-Net reconstruction can produce similar single-sided residuals. We retain these objects in our catalog, but we caution that their purity is significantly lower than that of Classes A and B. Higher-resolution imaging (e.g., HST (Hubble Space Telescope) or adaptive optics (AO)-assisted ground-based observations) is necessary to resolve their true nature.
\end{description}
Finally, the bottom row of Fig. \ref{fig:candidates} presents the "false positives" rejected during visual inspection, highlighting the limitations of the networks. Despite their high probability scores ($P_{\mathrm{ResNet}} > 0.6$), these objects were identified by human experts as contaminants because they exhibit features unrelated to lenses. 

This hierarchical classification strategy results in a final catalog of 88 Class A, 322 Class B, and 1,758 Class C candidates. This graded sample serves as a flexible resource for future studies, allowing researchers to choose between high purity (Class A) and high completeness (Classes A, B, and C) based on their scientific goals. We then cross-matched our candidates with the Master Lens Database, a comprehensive catalog of approximately 11,629 lens candidates \footnote{The candidates are available at \url{https://test.masterlens.org/}.}.
Our matching results yield ten Class A, six Class B, and 29 Class C candidates from our sample, totaling 45 lenses, displayed in Fig. \ref{fig:match} and ~\ref{fig:match_apd}. Where available, we also incorporated the grades assigned in the original studies (\citealt{2008ApJ...682..964B, 2018PASJ...70S..29S, 2019MNRAS.484.3879P}), except for a limited number of samples where we could not locate the matching grades in the original study. We aligned their categorical ratings (\textit{Sure Lens}, \textit{Maybe Lens} and \textit{No Lens}) with our Class A, B, and C categories, respectively. Notably, the grading levels assigned in these previous studies were generally lower than those achieved in our study. This discrepancy is largely attributable to the advantages provided by lens-subtracted images in our analysis. For many candidates, particularly in Classes B and C, arcs were not readily apparent until after foreground light subtraction, which revealed their presence.

\section{Discussion}
\label{sec:discussion}
In this section, we first dive deeper into the design of our training dataset and the distribution of the U-Net through comparative test. We then turn to discuss the core systematic limitations of our current detection framework. Finally, we discuss the broader scientific implications of our method and how this residual-based detection strategy can be adapted to support upcoming large-scale imaging surveys.

\begin{figure}[t!]
\centering
\includegraphics[width=\hsize]{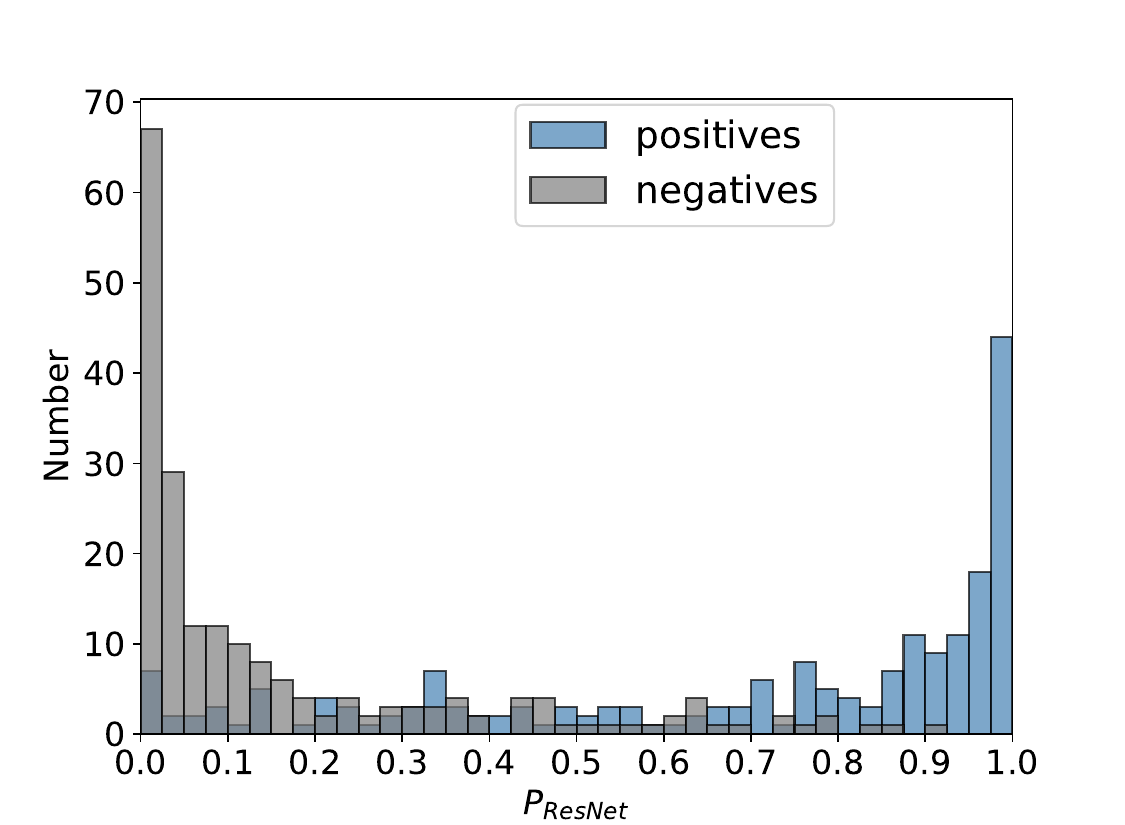}
    \caption{ResNet testing results of balanced strategy. Except for the 1:1 positive-to-negative sample ratio of the training dataset, all settings match those in the middle subfigure in Fig. \ref{fig:testresult}.}
    \label{fig:traindistribution}
\end{figure}

\subsection{Training strategies and limitations}
The efficacy of deep learning in astronomical object detection is highly dependent on the training set. In this work, we employed an unbalanced training dataset during the foreground modeling phase. While standard machine learning paradigms typically use balanced datasets to avoid class bias, our experiments demonstrate that for reconstructing foreground light from strong lensing— a rare phenomenon in the universe— a negatively dominant distribution (with a 1:10 positive-to-negative ratio) yields better overall performance. 

\begin{figure*}
    \centering
    \includegraphics[width=0.98 \textwidth]{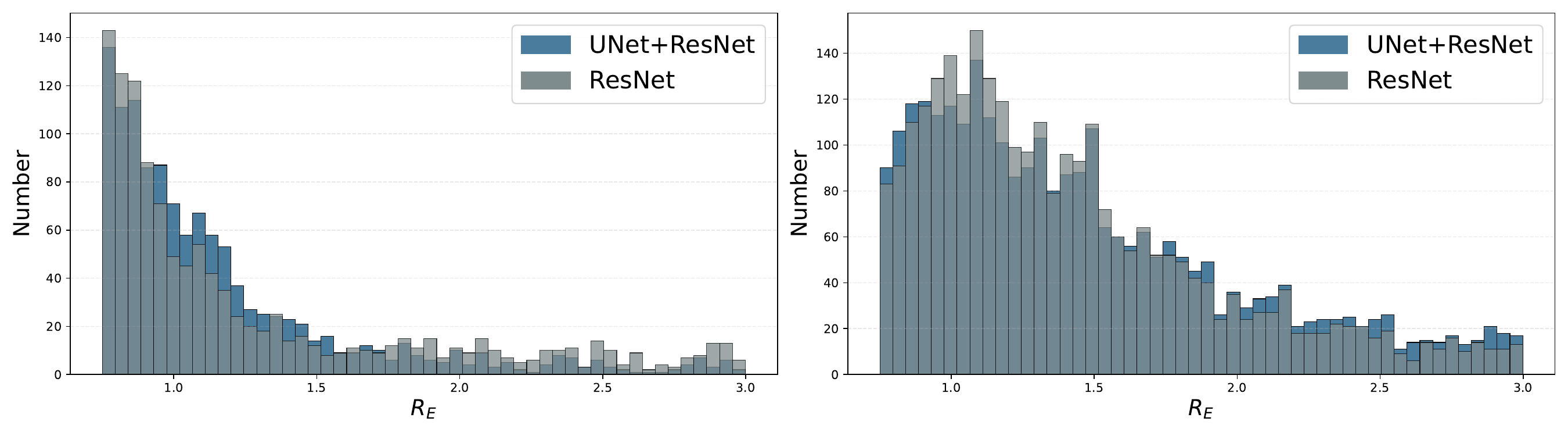}
    \caption{Comparative analysis of lens detection performance between the U-Net plus ResNet and ResNet-only models, illustrating the distribution of Einstein radii ($R_{\mathrm{E}}$) within different confidence segments of their predictions. Left: $R_{\mathrm{E}}$ distribution for predictions falling within the lower 30\% confidence range (low-confidence detections). Right: the $R_{\mathrm{E}}$ distribution for predictions within the upper 70\% confidence range (high-confidence detections). This figure quantifies the impact of U-Net integration on lens classification across varying $R_{\mathrm{E}}$.}
    \label{fig:u_r_test}%
\end{figure*}

We evaluated two training strategies for U-Net with different positive-to-negative ratios: balanced ($1:1$) (Fig. \ref{fig:traindistribution}) and negative-dominant ($1:10$) (the middle panel of Fig. \ref{fig:testresult}). Apart from the dataset composition, the training procedures are identical. The pipeline involved first training the U-Net to subtract foreground light, then using the residuals to train the ResNet for final prediction on the real testing data (\texttt{Test\_real}). The probability distributions for the balanced and negative-dominant data trained models are shown in Fig. \ref{fig:testresult} and Fig. \ref{fig:traindistribution}, respectively. Initial observations indicate that the balanced strategy classifies more negative samples as positives than the negative-dominant strategy, resulting in a higher FPR. For example, at a probability threshold of $0.6$, the balanced strategy yields an FPR of $8\%$, which is higher than the $4.5\%$ observed in the negative-dominant results. Furthermore, at the same threshold, the negative-dominant strategy achieves a higher TPR of $71.5\%$, compared to $68\%$ for the balanced strategy. We further find that when fixing the TPR at $0.7$, the negative-dominant strategy achieves a lower FPR of $4\%$ (classified by ResNet). In comparison, the balanced strategy results in an FPR of $9\%$. Overall, the negative-dominant strategy outperforms the balanced approach across varying probability thresholds. This performance difference illustrates distinct tendencies in the feature extraction and classification process, likely influenced by the rarity of the target class. Specifically, the balanced strategy assigns higher probability scores to any target exhibiting lens-like features, thereby increasing scores for non-lens instances. In contrast, the negative-dominant strategy yields more definitive classifications with superior TPR and FPR metrics. This reduction in FPs significantly facilitates our visual inspection process and is a primary justification for selecting the negative-dominant strategy.

We attribute these differences to the role of U-Net as a feature extractor with a prior. When trained on a balanced dataset, the network learns the prior information that "lensing is common," which may cause it to misclassify false lensed arcs to minimize reconstruction loss. The 1:10 ratio imposes a strong prior that non-lensed objects dominate the sky. This compels the U-Net to develop a robust and conservative representation of the foreground galaxy type. Consequently, we adopted the negative-majority distribution to prevent the candidate list from being overwhelmed by contaminants during subsequent visual inspections. It is crucial to distinguish the roles of the reconstruction network and the classification network. The U-Net acts as a "conservative filter," designed to suppress background noise from standard galaxies, benefiting from the 1:10 imbalance. In contrast, the ResNet functions as the "unbiased judge," and is strictly trained on a balanced (1:1) dataset.  This decoupling ensures that the decision boundary of the final classifier remains objective and is not shifted by the rarity of the target class.

Despite the feasibility of the pipeline, the methodology is subject to several limitations stemming from both the training data generation and the observational constraints of the KiDS survey. The most significant uncertainty arises from the simplification of galaxy mass profiles. Our simulation relies on the SIE model for deflectors and S\'ersic profiles for both foreground light and lensing sources. While these analytic profiles provide a first-order approximation of gravitational potentials and light, real galaxies exhibit a rich diversity of higher-order morphological complexities, including bright knots, boxy or disky deviations, arms or rings. A U-Net trained primarily on smooth SIE and S\'ersic profiles may misinterpret these intrinsic galactic irregularities as "residuals" after subtraction. For example, the spiral arms of a face-on late-type galaxy, if not accurately modeled by the network, can mimic the curvature of a gravitational arc. This limitation is inherent to any simulation-based training approach where the generative model is less complex than the real universe. Furthermore, reliance on single-band ($r$-band) photometry imposes a fundamental physical limitation. In strong lensing scenarios, the deflector is typically a red, passive elliptical galaxy, while the source is a blue, star-forming galaxy. A single-band analysis sacrifices color information, making it challenging to distinguish between lensed arcs and the foreground galaxy.

Finally, our target selection prioritizes massive galaxies to maximize the lensing cross section, which inherently biases our search against lenses formed by lower-mass or late-type galaxies. Consequently, our results do not fully explore the lower end of the stellar mass function of lenses. Additionally, the sequential nature of our "subtract-then-classify" architecture introduces a dependency risk: ResNet performance is strictly limited by U-Net subtraction. If the U-Net fails to adequately remove a bright foreground core due to the aforementioned morphological mismatches, the ResNet cannot recover the true signal, as it operates solely on the contaminated residual map.

To assess the contribution and potential limitations of the introduction of U-Net, we conducted a comparative experiment. We trained a stand-alone ResNet classifier on the same dataset as our U-Net plus ResNet model, but without the foreground light subtraction step. This ResNet-only model was subsequently evaluated on the same test dataset. Given the inherent differences in prediction score distributions between network architectures, we adopted a comparative approach: we analyzed the percentage of detected lenses within the top 70\% (high-confidence) and bottom 30\% (low-confidence) segments of each model's output. Fig. \ref{fig:u_r_test} illustrates the performance across different $R_{\mathrm{E}}$. For smaller Einstein radii ($R_{\mathrm{E}} \lesssim 0.9 \arcsec$), the U-Net+ResNet architecture detects a greater percentage of lenses with higher confidence scores. Conversely, for larger Einstein radii ($0.9 \arcsec \lesssim R_{\mathrm{E}} \lesssim 1.5 \arcsec$), the stand-alone ResNet tends to identify a greater proportion of lenses. In this range, foreground contamination is generally less dominant, and the ResNet alone appears more sensitive to subtle lensing features, potentially by avoiding information loss or errors introduced during the U-Net's subtraction process. While the U-Net offers both advantages and disadvantages for lens identification across different $R_{\mathrm{E}}$ ranges, the outputs of the U-Net, such as the generated residual images, hold extensive research significance beyond a simple classification task. They are valuable for improving the efficiency and reliability of visual inspection campaigns, facilitating more accurate lens grading, and can benefit future lens modeling in gravitational lensing studies.

Future iterations of this pipeline will address current limitations by incorporating multiband inputs. A multichannel U-Net could learn to associate specific spectral signatures with the foreground model and others with potential arc residuals, effectively automating the color-selection techniques used by human inspectors. Additionally, employing domain adaptation techniques, such as transferring the style from real images to simulations, could bridge the gap between idealized SIE and S\'ersic profiles and the realistic diversity of the galaxy population. This approach would further enhance the purity and completeness of future strong lens catalogs.

\subsection{Scientific implications and future prospects}
The methodology developed in this work offers a novel pathway to probe specific and scientifically valuable regions of strong lenses that are hidden foreground light. These lenses may possess a small $R_{\mathrm{E}}$ or have intrinsically faint lensed images. For such lenses, the deflector could be less massive than most known lenses or located at extremely low or high redshifts. Identifying these lenses could extend the parameter space of strong lenses. For example, the detection of small Einstein radius lenses is particularly important for constraining the density profiles of galaxies at small radii. While most known lenses are massive ellipticals with large $R_{\mathrm{E}}$ that probe the mass distribution where dark matter begins to dominate, these lenses investigate the inner regions where the interplay between baryonic matter and dark matter is most complex. By discovering "hidden" lenses, where the arc is obscured by the deflector's glare, we provide a sample that can help study low- and very high-redshift galaxies.

Looking ahead, the "foreground subtraction" paradigm established here is particularly advantageous for upcoming high-resolution and deep surveys. The method is ideally suited for space-based missions such as \textit{Euclid} and CSST, as well as deep ground-based projects such as LSST. In space missions, the stable, diffraction-limited PSF will significantly enhance the efficacy of the U-Net subtraction, preserving the sharp contrast between compact arcs and smooth galaxy halos and enabling the detection of arcs at separations as small as $0.2''-0.3''$. Conversely, for deep ground-based surveys such as LSST, while atmospheric seeing remains a challenge, the unprecedented depth will allow our method to model and remove extended deflector light, revealing low-surface-brightness arcs that are currently undetectable.

\section{Conclusion}
\label{sec:conclusion}
To address the limitations of previous lens search efforts, particularly for samples with smaller $R_{\mathrm{E}}$ or faint lensed images, we developed a composite CNN framework that utilizes both U-Net and ResNet architectures for feature extraction and classification. We trained our models using simulated data and evaluated their performance against both simulated and real datasets of lensing and non-lensing cases. The mock training data were simulated using an SIE mass profile combined with S\'ersic light profiles. The ray-tracing lensing equation was solved to generate the lensed images. We then convoluted the simulated images with the local PSFs corresponding to their respective coordinates and injected these into real observed cutouts. Additionally, we established a relationship between light and mass in the simulation using parameters such as photometric redshifts and S\'ersic parameters to enhance realism. Subsequently, we developed and trained two neural network architectures, U-Net and ResNet, to perform lens feature extraction and classification tasks, respectively. The U-Net model was used to isolate foreground light from full galaxy images, while the ResNet model classified these residuals to detect lens arcs. The trained models were applied to 638,398 galaxies selected from the KIDS-DR4 dataset, yielding 2,168 high-probability samples that were visually verified. Among these, there are 88 Class A, 322 Class B, and 1,758 Class C candidates. This approach can be extended to Stage 4 surveys, including the Euclid mission, CSST, and LSST.

\section{Data availability}
The visual classification catalog along with the match catalog are  available in electronic form at the CDS via anonymous ftp to \url{cdsarc.u-strasbg.fr (130.79.128.5)} or via \url{http://cdsweb.u-strasbg.fr/cgi-bin/qcat?J/A+A/}.

\begin{acknowledgements}
    This work is supported by the National Natural Science Foundation of China (Grant No. 12588202) and the China Manned Space Program with grant No.CMS-CSST-2025-A03. Rui Li acknowledges the National Natural Science Foundation of China (No. 12203050) and the Natural Science Foundation of Henan Province of China (Grant No. 252300423008).
\end{acknowledgements}

\bibliographystyle{bibtex/aa.bst} 
\bibliography{bibtex/cite.bib} 

\onecolumn
\appendix
\section{Additional figures}
We present additional examples of the U-Net prediction results in Fig.~\ref{fig:u_pred_real_apd}, as a continuation of Fig.~\ref{fig:u_pred_real}. We also show the complete set of our final candidates that were matched with the 11,629 lens candidates in the Master Lens Database, as described in Sect.~\ref{subsec:visual_classification}, in Fig.~\ref{fig:match_apd}, as a continuation of Fig.~\ref{fig:match}.

\begin{figure*}[!ht] 
    \centering
    \begin{minipage}{0.46\textwidth}
        \centering
        \includegraphics[width=\linewidth]{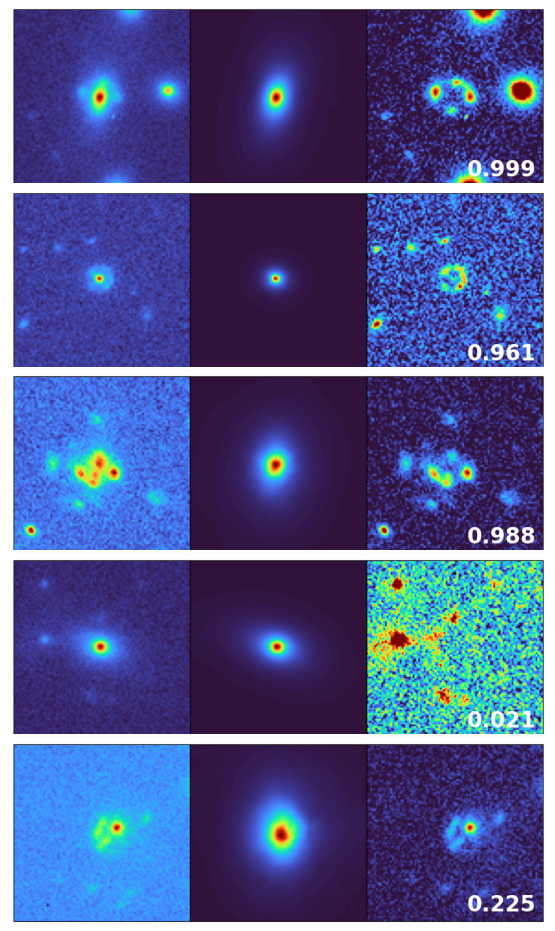}
    \end{minipage}
    \begin{minipage}{0.46\textwidth}
        \centering
        \includegraphics[width=\linewidth]{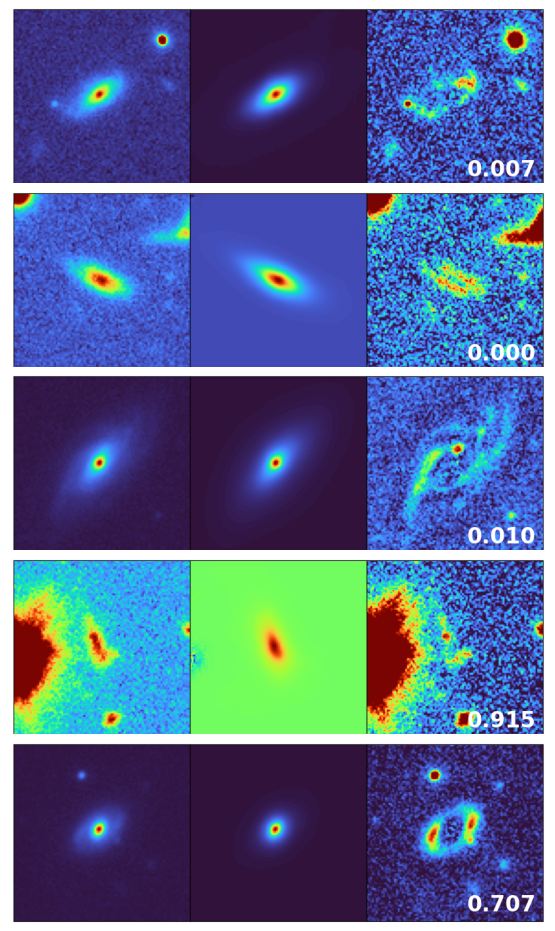}
    \end{minipage}
    \caption{Continuation of Fig. \ref{fig:u_pred_real}.}
    \label{fig:u_pred_real_apd}
\end{figure*}
\vspace{-5mm}

\begin{figure}[!h] 
    \centering
    \includegraphics[width=0.97\textwidth]{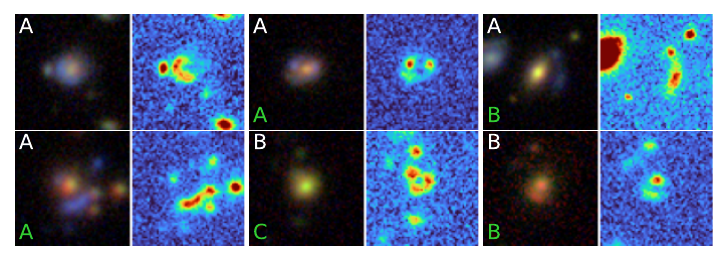}
    \caption{Continuation of Fig. \ref{fig:match}.}
    \label{fig:match_apd}
\end{figure}

\begin{figure}[!htbp] 
    \centering
    \includegraphics[width=0.97\textwidth]{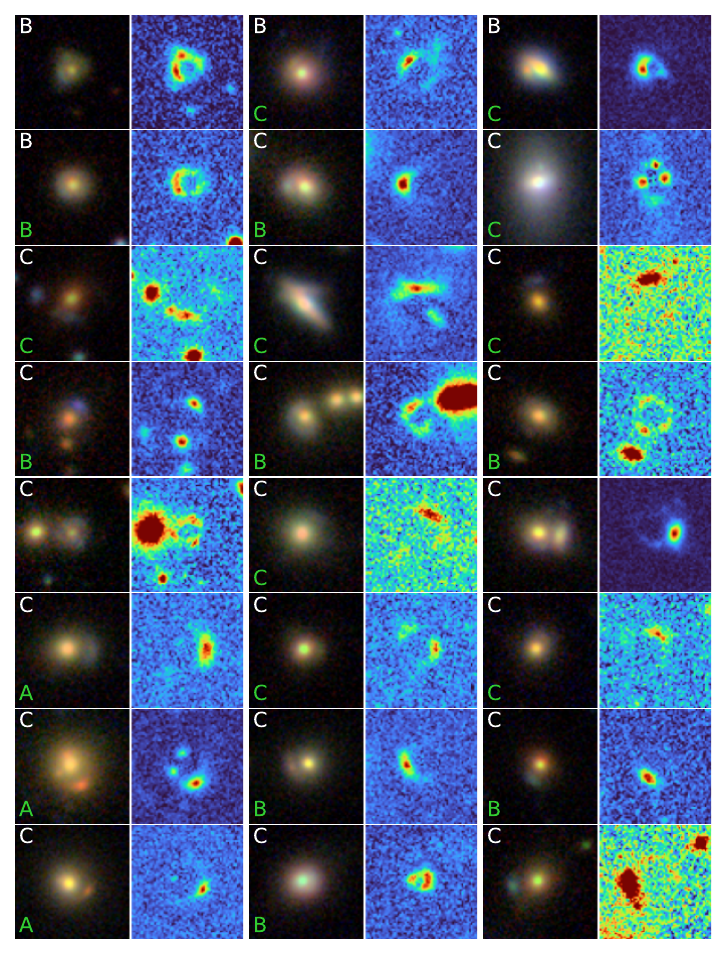}
    \caption*{Fig. A.2:(Continued)}
\end{figure}

\begin{figure}[!htbp]
    \centering
    \includegraphics[width=0.97\textwidth]{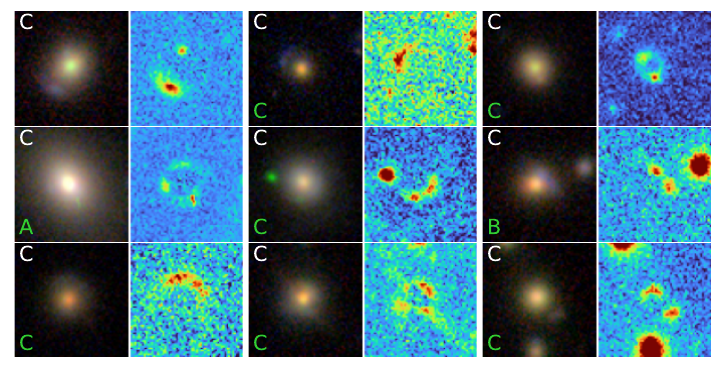}
    \caption*{Fig. A.2:(Continued)}
\end{figure}

\end{document}